\documentclass[11pt,superscriptaddress]{revtex4-1}

\usepackage{amsmath,bbm}
\usepackage{amssymb}
\usepackage{graphicx}
\usepackage{mathrsfs}
\usepackage{color}

\newcommand{\Tr}{\textrm{Tr}}
\newcommand{\ket}[2][]{{|#2\rangle_{#1}}}
\newcommand{\bra}[2][]{{}_{#1}\langle #2|}
\newcommand{\braket}[3][]{{{}_{#1}\langle#2|#3\rangle_{#1}}}
\newcommand{\proj}[2][]{\ket{#2}_{#1}\bra{#2}}

\newcommand{\ketbra}[2]{\ensuremath{\ket{#1} \bra{#2}}}
\newcommand{\abs}[1]{\ensuremath{\lvert #1 \rvert}}

\begin{document}

\title{Low-cost limit of classical communication with restricted quantum measurements}

\author{Ludwig Kunz}
\email{Corresponding author: l.kunz@cent.uw.edu.pl}
\affiliation{Centre for Quantum Optical Technologies, Centre of New Technologies, University of Warsaw, Banacha 2c, 02-097 Warszawa, Poland}
\affiliation{Faculty of Physics, University of Warsaw, Pasteura 5, 02-093 Warszawa, Poland}

\author{Marcin Jarzyna}
\affiliation{Centre for Quantum Optical Technologies, Centre of New Technologies, University of Warsaw, Banacha 2c, 02-097 Warszawa, Poland}

\author{Wojciech Zwoli\'{n}ski}
\affiliation{Centre for Quantum Optical Technologies, Centre of New Technologies, University of Warsaw, Banacha 2c, 02-097 Warszawa, Poland}
\affiliation{Faculty of Physics, University of Warsaw, Pasteura 5, 02-093 Warszawa, Poland}

\author{Konrad Banaszek}
\affiliation{Centre for Quantum Optical Technologies, Centre of New Technologies, University of Warsaw, Banacha 2c, 02-097 Warszawa, Poland}
\affiliation{Faculty of Physics, University of Warsaw, Pasteura 5, 02-093 Warszawa, Poland}

\begin{abstract}
We consider a communication scenario where classical information is encoded in an ensemble of quantum states that admit a power series expansion in a cost parameter and with the vanishing cost converge to a single zero-cost state. For a given measurement scheme, we derive an approximate expression for mutual information in the leading order of the cost parameter. The general results are applied to selected problems in optical communication, where coherent states of light are used as input symbols and the cost is quantified as the average number of photons per symbol. We show that for an arbitrary individual measurement on  phase shift keyed (PSK) symbols, the photon information efficiency is upper bounded by 2 nats of information per photon in the low-cost limit, which coincides with the conventional homodyne detection bound. The presented low-cost approximation facilitates a systematic analysis of few-symbol measurements that exhibit superadditivity of accessible information. For the binary PSK alphabet of coherent states, we present designs for two- and three-symbol measurement schemes based on linear optics, homodyning, and single photon detection that offer respectively 2.49\% and 3.40\% enhancement relative to individual measurements. We also show how designs for scalable superadditive measurement schemes emerge from the introduced low-cost formalism.
\end{abstract}

\maketitle

\section{Introduction}

The celebrated Holevo quantity \cite{Holevo1973} provides a powerful tool to identify how much classical information can be transmitted using symbols drawn from a given ensemble of quantum states. It specifies an upper bound on the mutual information that can be achieved with any quantum measurement satisfying physical constraints. In general, attaining the Holevo quantity requires a collective measurement on an arbitrary number of symbols \cite{Hausladen1996, Holevo1998,GuhaWildeISIT2012,WildeGuhaISIT2012}.
When the available class of measurements is restricted, for example to individual or few-symbol detection, optimization of the mutual information with respect to the measurement strategy often becomes a rather challenging task for which universal methods are missing. Interestingly, the maximum mutual information that can be obtained using measurements on a restricted number of symbols, known as accessible information, exhibits superadditive behavior with the number of symbols \cite{Holevo1979, Peres1991}.
This is intimately related to the fact that quantum measurement reveals in general only partial knowledge about the measured states and more relevant information can be retrieved through collective detection strategies. Theoretical analysis of superadditivity of accessible information is highly nontrivial even for binary ensembles of elementary input states \cite{Sasaki1998,Buck1999,Guha2011}. Another problem is the actual physical implementation of optimal measurements, e.g.\ in the optical domain.

The purpose of this paper is to present a systematic expansion of the mutual information in a cost parameter which characterizes the ensemble of quantum states used for communication. It is assumed that the quantum states admit a power series expansion in the cost parameter and converge to the same zero-cost state when the value of the cost parameter tends to zero. This allows us to derive an asymptotic bound on the mutual information in the leading order of the cost parameter for a given measurement scheme. An essential advantage of the formalism developed here is that the characteristics of the ensemble enters the asymptotic expression through a set of state vectors that are independent of the cost parameter. As we demonstrate here, this greatly simplifies identification of bounds on mutual information attainable with few-symbol measurements and provides a systematic method to construct experimental schemes that demonstrate superadditivity of accessible information.

The usefulness of the presented approach is illustrated with a number of examples motivated by optical communication. In this context, a natural measure for the communication cost is the mean photon number per symbol. Our expansion provides non-trivial bounds on the photon information efficiency (PIE), which specifies the amount of information that can be transmitted in one photon. In particular, we show that when information is encoded in the phase of coherent states, the maximum attainable PIE is 2~nats per photon in the low-cost limit when arbitrary individual measurements are permitted. This coincides with the commonly used in optical communications reference value derived from the Shannon-Hartley theorem, which assumes the specific case of homodyne detection
\cite{BanaszekJLT2020}. We also provide a relation between the attainable PIE and the peak-to-average power ratio of a general coherent state ensemble used for communication. Further, we investigate superadditivity of accessible information in the case of few-symbol measurements performed on the binary ensemble of coherent states with equal mean photon number and opposite phases. We show that linear optical setups combining photon counting and homodyne detection can be used to demonstrate superadditivity for coherent states carrying on average less than approx.\ $0.01$ photon. We also discuss how scalable communication schemes exhibiting the superadditivity effect \cite{Guha2011, Banaszek2017} emerge from the presented formalism.

This paper is organized as follows. Sec.~\ref{sec:problem} presents the mathematical formulation of the problem. Asymptotic expansion of mutual information in the limit of the vanishing cost is carried out in Sec.~\ref{sec:expansion}.
The results are applied to communication with an ensemble of coherent states and individual measurements in
Sec.~\ref{Sec:Individual}. The case of few-symbol measurements on the binary ensemble of coherent states is discussed in Sec.~\ref{sec:collective}. Finally, Sec.~\ref{sec:conclusions} concludes the paper.

\section{Problem formulation}
\label{sec:problem}

A communication scheme can be regarded as encoding messages onto quantum states of certain physical carriers which after transmission are subsequently measured at the receiver. The measurement results are processed to decode the input message.
We shall consider classical information encoded in a finite ensemble of quantum states, labeled using an index $j$ and described, after transmission, by density operators $\hat{\varrho}_j(\zeta)$, with respective input probabilities $p_j$. The states depend on a real nonnegative parameter $\zeta$ characterizing their cost.
Physically, the cost may correspond e.g.\ to the mean photon number per symbol in the case of optical communication.
We will assume that each $\hat{\varrho}_j(\zeta)$ is an analytical operator-valued function of $\zeta$ and that in the limit $\zeta \rightarrow 0$ all the states $\hat{\varrho}_j(\zeta)$ converge to the same state $\hat{\varrho}^{(0)}$ which we shall call the {\em zero-cost state}. Furthermore, we will assume that the states  $\hat{\varrho}_j(\zeta)$ admit a power series expansion in the cost parameter around $\zeta=0$. The asymptotic analysis will be based on an expansion up to the second order:
\begin{equation}
\label{Eq:rhoexp}
\hat{\varrho}_j(\zeta) \approx \hat{\varrho}^{(0)} + \zeta \hat{\varrho}^{(1)}_j + \zeta^2 \hat{\varrho}^{(2)}_j ,
\end{equation}
where
\begin{equation}
\label{Eq:rhoexpansionterms}
\hat{\varrho}^{(1)}_j = \left. \frac{\partial\hat{\varrho}_j}{\partial\zeta}\right|_{\zeta=0}, \qquad
\hat{\varrho}^{(2)}_j = \left.\frac{1}{2}\frac{\partial^2\hat{\varrho}_j}{\partial\zeta^2}\right|_{\zeta=0}.
\end{equation}
Note that the zeroth-order term is independent of $j$ as it is given by the zero-cost state. The above expansion is assumed to be valid for sufficiently small values of the cost parameter $\zeta$.

The most general measurement that can be performed on a
quantum system is described by a positive operator-valued measure (POVM) with elements $\hat{Q}_r$ that are positive semidefinite and add up to the identity operator, $\sum_{r} \hat{Q}_r = \hat{\openone}$. The index $r$ specifies the measurement result. For simplicity we consider here a finite set of possible measurement outcomes.
The conditional probability $p_{r|j}(\zeta)$ of obtaining an outcome $r$ given the $j$th input state is defined by Born's rule
\begin{equation}
\label{Eq:Born's}
p_{r|j}(\zeta) = \Tr [\hat{Q}_r \hat{\varrho}_j(\zeta) ].
\end{equation}
The marginal probability of the $r$th outcome for the ensemble is $p_r(\zeta) = \sum_{j} p_j p_{r|j}(\zeta)$.

The maximum amount of classical information that can be transmitted using the input ensemble of quantum states $\hat{\varrho}_j$ with respective probabilities $p_j$ and a measurement $\hat{Q}_r$ is given by mutual information ${\sf I}$ calculated for the joint probability distribution $p_j p_{r|j}(\zeta)$, where the conditional probabilities $p_{r|j}(\zeta)$ are given by Eq.~(\ref{Eq:Born's}). It will be convenient to decompose mutual information into a sum of contributions from individual measurement results,
\begin{equation}\label{eq:I}
{\sf I}=\sum_{r} {\sf I}_r
\end{equation}
 with ${\sf I}_r$ given by
\begin{equation}
\label{Eq:Ir}
 {\sf I}_r =  \sum_{j} p_j p_{r|j}(\zeta) \log \left( \frac{p_{r|j}(\zeta)}{p_{r}(\zeta)}\right).
\end{equation}
Throughout this paper we use natural logarithms and the information is measured in nats, $1~{\textrm{nat}} = \log_2 e~\textrm{bits}$.

Our overall objective will be to optimize mutual information over measurement operators chosen from a restricted class and input probabilities for a given set (constellation) of symbols $\varrho_j(\zeta)$ in the regime of a small cost parameter. The first step, presented in the next section, will be a systematic expansion of the mutual information ${\sf I}$ in the cost parameter $\zeta$ that will provide an upper bound on the leading-order term. From now on, it is assumed that the measurement operators $\hat{Q}_r$ do not depend on the cost parameter. This restricted scenario is motivated by practical optical communication that uses direct detection or homodyne detection. We will see that even under this restriction the low-cost analysis of mutual information yields non-trivial results regarding the superadditivity of accessible information. In Sec.~\ref{Subsec:2copy} we will present an example of further enhancement when the measurement can be adjusted to the value of the cost parameter.

\section{Asymptotic expansion}
\label{sec:expansion}

In the limit of the vanishing cost parameter, Eq.~(\ref{Eq:Ir}) can be expressed through expansion of the  conditional probabilities $p_{r|j}(\zeta)$ in $\zeta$, which also determines marginal probabilities $p_r(\zeta)$.  As for $\zeta \rightarrow 0$ all the states $\hat{\varrho}_j (\zeta)$ converge to the zero-cost state $\hat{\varrho}^{(0)}$, it will be convenient to introduce a shorthand notation
\begin{equation}\label{eq:p0}
p_{r}^{(0)} = \Tr [\hat{Q}_r \hat{\varrho}^{(0)}] =  p_{r|j}(0) = p_{r} (0) .
\end{equation}
The expansion given in Eq.~(\ref{Eq:rhoexp}) implies the following approximate expressions for conditional probabilities
\begin{equation}
p_{r|j}(\zeta) \approx p_{r}^{(0)} + \zeta p_{r|j}^{(1)} + \zeta^2 p_{r|j}^{(2)}
\label{Eq:prj(zeta)}
\end{equation}
where $p_{r|j}^{(k)} = \Tr[\hat{Q}_r \hat{\varrho}^{(k)}_j]$, $k=1,2$.
The marginal probabilities are correspondingly approximated by
\begin{equation}
p_{r}(\zeta) \approx p_{r}^{(0)} + \zeta p_{r}^{(1)} + \zeta^2 p_{r}^{(2)}
\label{Eq:pr(zeta)}
\end{equation}
with $p_{r}^{(k)} = \sum_{j} p_j p_{r|j}^{(k)}$, where $k=1,2$. Equivalently, one
can write $p_{r}^{(k)} = \Tr[\hat{Q}_r \hat{\varrho}^{(k)}_{\textrm{ens}}]$, where the operators $\hat{\varrho}^{(k)}_{\textrm{ens}}$ are given by convex combinations
\begin{equation}
\hat{\varrho}^{(k)}_{\textrm{ens}} = \sum_{j}p_j \hat{\varrho}^{(k)}_{j}, \qquad k=1,2
\end{equation}
of derivatives defined in Eq.~(\ref{Eq:rhoexpansionterms}).
It should be noted that the zeroth-order term for both conditional probabilities in Eq.~(\ref{Eq:prj(zeta)}) and marginal probabilities in Eq.~(\ref{Eq:pr(zeta)}) is identical.

Further steps for a given measurement operator $\hat{Q}_r$ depend on whether the outcome $r$ can be generated by the zero-cost state or not. In the former case, corresponding to the mathematical condition
$p_{r}^{(0)} = \Tr [\hat{Q}_r \hat{\varrho}^{(0)} ] >0$,
we will call $\hat{Q}_r$ a {\em type-$\mathscr{Z}$} operator and use $\mathscr{Z}$ to denote the corresponding set of measurement outcomes. Measurement operators for which $p_{r}^{(0)} = \Tr [\hat{Q}_r \hat{\varrho}^{(0)} ] =0$ will be referred to as {\em type-$\mathscr{Z}^\perp$} operators and the set of measurement outcomes produced by these operators will be denoted by $\mathscr{Z}^\perp$. We will decompose
\begin{equation}
\label{Eq:IinZinZperp}
{\sf I} = \sum_{r \in \mathscr{Z}} {\sf I}_r + \sum_{r \in \mathscr{Z}^\perp} {\sf I}_r
\end{equation}
and analyze separately contributions to mutual information from the two sets of measurement outcomes.

In the case of type-$\mathscr{Z}$ operators, the logarithm in Eq.~(\ref{Eq:Ir}) can be expanded into a power series in the cost parameter $\zeta$, which yields in the leading order
\begin{equation}
\label{Eq:tildeIrdef}
{\sf I}_r \approx \tilde{\sf I}_r = \frac{\zeta^2}{2 p_r^{(0)}}
 \sum_j p_j \left(p_{r|j}^{(1)} - p_{r}^{(1)} \right)^2  .
\end{equation}
The difference $p_{r|j}^{(1)} - p_{r}^{(1)}$ appearing under summation can be written as
\begin{equation}
\label{Eq:prj1-pr1}
p_{r|j}^{(1)} - p_{r}^{(1)} = \Tr [\hat{Q}_r (\hat{\varrho}^{(1)}_j - \hat{\varrho}^{(1)}_\textrm{ens})].
\end{equation}
We will transform this expression further using techniques borrowed from quantum estimation theory \cite{Helstrom1976, Holevo1982, Braunstein1994, BarndorffNielsen2000}. Let us recall the notion of the symmetric logarithmic derivative (SLD), defined implicitly at $\zeta=0$ for individual and ensemble density operators by respective equations
\begin{align}
\label{Eq:SLDj}
\hat{\varrho}_{j}^{(1)} & = \left. \frac{\partial\hat\varrho_j}{\partial\zeta} \right|_{\zeta=0} = \frac{1}{2} \bigl( \hat{L}_j \hat{\varrho}^{(0)}
+ \hat{\varrho}^{(0)} \hat{L}_j \bigr), \\
\label{Eq:SLDens}
\hat{\varrho}_{\textrm{ens}}^{(1)} & = \left. \frac{\partial\hat\varrho_\textrm{ens}}{\partial\zeta} \right|_{\zeta=0} = \frac{1}{2} \bigl( \hat{L}_\textrm{ens} \hat{\varrho}^{(0)}
+ \hat{\varrho}^{(0)} \hat{L}_\textrm{ens} \bigr),
\end{align}
that are assumed to be well defined \cite{Liu2016,Safranek2017} for physical scenarios considered here.
Subtracting the expression in Eq.~(\ref{Eq:SLDens}) from Eq.~(\ref{Eq:SLDj}) and inserting the difference into Eq.~(\ref{Eq:prj1-pr1}) yields
\begin{equation}
p_{r|j}^{(1)} - p_{r}^{(1)}  = \frac{1}{2} \Tr [\hat{Q}_r ( \hat{D}_j \hat{\varrho}^{(0)} + \hat{\varrho}^{(0)}
\hat{D}_j)],
\end{equation}
where
\begin{equation}
\label{Eq:Ldef}
\hat{D}_j = \hat{L}_j - \hat{L}_\textrm{ens}.
\end{equation}
Using the Schwarz inequality
\begin{equation}
|\Tr [\hat{A} \hat{B}]|^2 \le \Tr [\hat{A}^\dagger \hat{A}] \Tr [\hat{B}^\dagger \hat{B}]
\end{equation}
applied to the operators
\begin{equation}
\hat{A} = \frac{1}{\sqrt{p_r^{(0)}}}
\sqrt{\hat{\varrho}^{(0)}} \sqrt{\hat{Q}_r} , \quad
\hat{B} = \sqrt{\hat{Q}_r} \hat{D}_j \sqrt{\hat{\varrho}^{(0)}}
\end{equation}
and their hermitian conjugates $\hat{A}^\dagger$ and $\hat{B}^\dagger$ and inserting the result into
Eq.~(\ref{Eq:tildeIrdef})
yields the following upper bound on the mutual information contribution from the measurement outcome $r \in \mathscr{Z}$:
\begin{equation}
\label{Eq:Ir<=SLD}
\tilde{\sf I}_r \le \frac{\zeta^2}{2} \Tr \left[ \hat{Q}_r \left( \sum_j p_j  \hat{D}_j \hat{\varrho}^{(0)} \hat{D}_j \right) \right].
\end{equation}
Note that the obtained bound is quadratic in the cost parameter $\zeta$.

In the case of type-$\mathscr{Z}^\perp$ measurement operators, the condition $\Tr[\hat{Q}_r \hat{\varrho}^{(0)}] = 0$ combined with positive semidefiniteness of the operators $\hat{Q}_r$ and $\hat{\varrho}^{(0)}$ implies that $\hat{Q}_r \hat{\varrho}^{(0)} = \hat{\varrho}^{(0)}\hat{Q}_r = 0$.
Therefore, a calculation of $\Tr[\hat{Q}_r \hat{\varrho}_j^{(1)}]$ using the SLD defined in Eq.~(\ref{Eq:SLDj}) yields zero and the leading-order term in the power series expansion of the conditional probability $p_{r|j}(\zeta) = \Tr[\hat{Q}_r \hat{\varrho}_j(\zeta)]$ is quadratic in $\zeta$. We will write it in the form
\begin{equation}
p_{r|j}(\zeta) \approx \zeta^2 \Tr[\hat{Q}_r \hat{\varrho}'^{(2)}_j],
\end{equation}
where
\begin{equation}
\label{Eq:varrhoprim(2)}
\hat{\varrho}'^{(2)}_j = \hat{\Pi} \hat{\varrho}^{(2)}_j \hat{\Pi}
\end{equation}
with $\hat{\varrho}^{(2)}_j$ defined in Eq.~(\ref{Eq:rhoexpansionterms}) and $\hat{\Pi}$ being the projection onto the kernel of the zero-cost state $\hat{\varrho}^{(0)}$. This follows from the fact that for type-$\mathscr{Z}^\perp$ operators one has $\hat{\Pi} \hat{Q}_r \hat{\Pi} = \hat{Q}_r$.

As a result, the contribution to mutual information generated by a measurement outcome $r \in \mathscr{Z}^\perp$ can be written in the form:
\begin{equation}
\label{Eq:Irno0}
{\sf I}_r \approx \tilde{\sf I}_r = \zeta^2 \sum_j p_j \Tr\bigl[\hat{Q}_r \hat{\varrho}'^{(2)}_j\bigr]
\log \frac{\Tr\bigl[\hat{Q}_r \hat{\varrho}'^{(2)}_j\bigr]}{\sum_{l} p_{l} \Tr\bigl[\hat{Q}_r \hat{\varrho}'^{(2)}_{l}\bigr]}.
\end{equation}
Interestingly, although type-$\mathscr{Z}^\perp$ operators generate events with vanishing probability in the limit $\zeta\rightarrow 0$, their contribution to mutual information is of the same order in $\zeta$ as from type-$\mathscr{Z}$ operators. A similar behavior was observed for mutual information in a somewhat different context in \cite{Czajkowski2017}.

Inserting Eqs.~(\ref{Eq:Ir<=SLD}) and (\ref{Eq:Irno0}) into Eq.~(\ref{Eq:IinZinZperp}), the leading-order expansion of mutual information in the cost parameter $\zeta$ is upper bounded by the following expression:
\begin{align}\label{eq:mutinf_expansZZperp}
{\sf I} \approx \tilde{\sf I} & \le \zeta^2 \sum_{r \in \mathscr{Z}^\perp} \sum_j p_j \Tr\bigl[\hat{Q}_r \hat{\varrho}'^{(2)}_j\bigr]
\log \frac{\Tr\bigl[\hat{Q}_r \hat{\varrho}'^{(2)}_j\bigr]}{\sum_{l} p_{l} \Tr\bigl[\hat{Q}_r \hat{\varrho}'^{(2)}_{l}\bigr]}
\nonumber \\
& + \frac{\zeta^2}{2} \Tr \left[ \left( \sum_{r \in \mathscr{Z}} \hat{Q}_r \right)\left( \sum_j p_j  \hat{D}_j \hat{\varrho}^{(0)} \hat{D}_j \right) \right].
\end{align}
The sum over $\mathscr{Z}$-type measurement operators appearing in the second line of the above formula can be expressed in terms of $\mathscr{Z}^\perp$-type operators as $\sum_{r \in \mathscr{Z}} \hat{Q}_r= \hat{\openone} - \sum_{r \in \mathscr{Z}^\perp} \hat{Q}_r$. This allows us to write the derived bound solely in terms of $\mathscr{Z}^\perp$-type operators as
\begin{align}\label{eq:mutinf_expans}
{\sf I} \approx \tilde{\sf I} & \le \zeta^2 \sum_{r \in \mathscr{Z}^\perp} \sum_j p_j \Tr\bigl[\hat{Q}_r \hat{\varrho}'^{(2)}_j\bigr]
\log \frac{\Tr\bigl[\hat{Q}_r \hat{\varrho}'^{(2)}_j\bigr]}{\sum_{l} p_{l} \Tr\bigl[\hat{Q}_r \hat{\varrho}'^{(2)}_{l}\bigr]}
\nonumber \\
& + \frac{\zeta^2}{2} \Tr \left[ \left( \hat{\openone} - \sum_{r \in \mathscr{Z}^\perp} \hat{Q}_r \right)\left( \sum_j p_j  \hat{D}_j \hat{\varrho}^{(0)} \hat{D}_j \right) \right].
\end{align}
As we will see in the following sections, the above expression, despite its seemingly complicated form, can help substantially in the analysis of specific communication scenarios in the low-cost limit.

\section{Individual measurements on single-mode coherent states}
\label{Sec:Individual}

The asymptotic analysis presented in Sec.~\ref{sec:expansion} can be applied to a variety of physical situations. In the following we will consider a scenario motivated by optical communication, when information is transmitted using an alphabet of coherent states of a single light mode \cite{Caves1994, Shapiro2009}
\begin{equation}
\ket{\zeta\gamma_j} = e^{- \zeta^2|\gamma_j|^2/2} \sum_{n=0}^{\infty} \frac{(\zeta\gamma_j)^n}{\sqrt{n!}} \ket{n}.
\label{Eq:CoherentStateAlphabet}
\end{equation}
Here $\ket{n}$ denotes the $n$-photon Fock state. The complex amplitudes $\zeta\gamma_j$ of coherent states are written as products of complex coefficients $\gamma_j$ and the nonnegative cost parameter $\zeta$ that can be viewed as the scaling factor for the entire ensemble represented as a constellation in the complex plane. Physically, it can describe the attenuation of an optical channel over which information transfer takes place. When $\zeta\rightarrow 0$, all the states converge to the vacuum state $\ket{0}$ which plays the role of the zero-cost state $\hat{\varrho}^{(0)} = \proj{0}$ introduced in Sec.~\ref{sec:problem}. If individual coherent states $\ket{\zeta\gamma_j}$ are used with respective probabilities $p_j$,
the average optical energy for the ensemble, measured in the photon number units, reads
\begin{equation}
\label{Eq:nbar=}
\bar{n} = \zeta^2 \sum_{j} p_j |\gamma_j|^2.
\end{equation}
and is proportional to the square of the cost parameter. As a simple example, $m$-ary phase shift keying (PSK) corresponds to taking $\gamma_j = e^{2\pi i j /m}$ and $p_j=1/m$ with $j=0,1,\ldots, m-1$. With this parametrization, all the input states have the optical energy equal to the average $\bar{n}=\zeta^2$.

In this section we consider a scenario when information is retrieved by measuring individually symbols in the form of single-mode coherent states defined in Eq.~(\ref{Eq:CoherentStateAlphabet}). It is straightforward to obtain explicit expressions for the two operators that enter the asymptotic expression for the mutual information given in Eq.~(\ref{eq:mutinf_expansZZperp}). The
SLD difference defined in Eq.~(\ref{Eq:Ldef}) reads
\begin{equation}
\label{Eq:Lcohens}
\hat{D}_j = 2[(\gamma_j -\gamma_{\textrm{ens}}) \ket{1}\bra{0} + (\gamma_j -\gamma_{\textrm{ens}})^\ast\ket{0}\bra{1}],
\end{equation}
where $\gamma_{\textrm{ens}} = \sum_j p_j \gamma_j$, while
\begin{equation}
\label{Eq:rho(2)=11}
\hat{\varrho}'^{(2)}_j = |\gamma_j|^2 \proj{1}.
\end{equation}
Consequently, it is sufficient to consider the action of measurement operators $\hat{Q}_r$ only in the two-dimensional subspace spanned by the zero- and the one-photon Fock states $\ket{0}$ and $\ket{1}$.

\subsection{Photon information efficiency bounds}
\label{Sec:PIEBounds}

Suppose first that the measurement is composed only of type-$\mathscr{Z}$ operators. Inserting Eq.~(\ref{Eq:Lcohens}) into Eq.~(\ref{Eq:Ir<=SLD}) and recalling that $\hat{\varrho}^{(0)} = \proj{0}$ yields for $r \in \mathscr{Z}$:
\begin{equation}
\label{Eq:Ir<=<1|Mr|1>(1)}
\tilde{\sf I}_r \le 2 \zeta^2 \bra{1} \hat{Q}_r \ket{1} \sum_j p_j |\gamma_j-\gamma_{\textrm{ens}}|^2.
\end{equation}
The sum over $j$ in the above formula gives the variance of the complex coefficients $\gamma_j$, which is upper bounded by the mean square $\sum_j p_j |\gamma_j|^2$. If all the measurement operators are type-$\mathscr{Z}$, one has $\sum_r \bra{1}\hat{Q}_r \ket{1} = 1$ which yields:
\begin{equation}
\label{Eq:Icohtype-Z}
\tilde{\sf I} \le 2 \zeta^2 \sum_j p_j |\gamma_j|^2 = 2 \bar{n},
\end{equation}
where in the second step we have used the calculation of the ensemble average photon number carried out in Eq.~(\ref{Eq:nbar=}). In optical communication, the ratio ${\sf I}/\bar{n}$ is known as the
{\em photon information efficiency} (PIE) specifying the maximum amount of information that can be carried by one received photon. Therefore Eq.~(\ref{Eq:Icohtype-Z}) can be viewed as an upper bound of $2$ nats per photon on the PIE in the asymptotic limit $\bar{n} \rightarrow 0$, which holds for measurements composed from type-$\mathscr{Z}$ operators $\hat{Q}_r$ satisfying the condition $\bra{0} \hat{Q}_r \ket{0} > 0$.

Interestingly, the asymptotic bound of $2$ nats per photon on the PIE in the limit $\bar{n} \rightarrow 0$ appears also in a scenario when information is encoded in a single quadrature of the optical mode measured by shot-noise limited homodyne detection \cite{BanaszekJLT2020}. In this case the so-called Shannon-Hartley limit for the mutual information reads \cite{Shannon1949, Verdu2002, Guo2005, Banaszek2019, Papen2019}
\begin{equation}\label{Eq:ShannonHartley}
{\sf I}_{\textrm{SH}} = \frac{1}{2} \log\left(1+4 \bar n\right)
\end{equation}
and is attained by using an input ensemble of coherent states with the Gaussian distribution for the measured quadrature and the conjugate quadrature set to zero. A semi-qualitative connection between this scenario and the bound derived in Eq.~(\ref{Eq:Icohtype-Z}) can be made as follows. As it is well known, decomposing POVM elements into positive rank-1 operators can make mutual information only equal or greater \cite{Davies1978, Sasaki1999}. Type-$\mathscr{Z}$ and rank-$1$ POVM elements $\hat{Q}_r$ with non-zero matrix elements $\bra{1} \hat{Q}_r \ket{1}$ which non-trivially contribute to Eq.~(\ref{Eq:Icohtype-Z}) must detect coherence between zero- and one-photon Fock states and consequently are sensitive to the phase of the input coherent state. Homodyne detection is an example of such a phase-sensitive measurement, although in the case of the homodyne POVM the classification criterion based on whether $\Tr[\hat{Q}_r\hat{\varrho}^{(0)}]=0$ or not used in Sec.~\ref{sec:expansion} becomes problematic. This is because the outcome of the homodyne measurement has the continuous form corresponding to the field quadrature which can take any real value and the conditional probabilities for homodyne POVM elements on the zero-cost vacuum state become arbitrarily small with the increasing absolute value of the quadrature. More generally, the low-cost limit of PIE attainable with individual phase-sensitive measurements coincides with that derived from the Shannon-Hartley theorem also when the input coherent states are transmitted over a phase-invariant Gaussian channel with excess noise, mapping them onto displaced thermal states. This case is discussed in in Appendix~\ref{Sec:thermalNoise}.

Let us now turn our attention to type-$\mathscr{Z}^\perp$ measurement operators $\hat{Q}_r$. The condition $\Tr[\hat{Q}_r \hat{\varrho}^{(0)}]= 0$ defining type-$\mathscr{Z}^\perp$ operators implies that in the zero- and one-photon subspace $\hat{Q}_r$ must be proportional to a projection $\proj{1}$ onto the one-photon state. Therefore in optical scenarios type-$\mathscr{Z}^\perp$ operators correspond to direct photon counting. Inserting Eq.~(\ref{Eq:rho(2)=11}) into Eq.~(\ref{Eq:Irno0}) yields the contribution to mutual information from a measurement outcome $r\in \mathscr{Z}^\perp$ in the form
\begin{equation}
\label{Eq:IrcohZperp}
\tilde{\sf I}_r = \zeta^2 \bra{1} \hat{Q}_r \ket{1} \sum_j p_j |\gamma_j|^2
\log \frac{|\gamma_j|^2}{\sum_{l} p_{l} |\gamma_{l}|^2}.
\end{equation}
A frequently encountered constraint in optical communication systems
is an upper limit on the peak-to-average power ratio ${\cal P}$ of the transmitted signal. In our case ${\cal P}$ is given by the ratio of the maximum mean photon number carried by one symbol to the average photon number in the input ensemble:
\begin{equation}
\label{Eq:PAPRdef}
{\cal P} = \frac{\max_j |\gamma_j|^2}{\sum_l p_l |\gamma_l|^2}.
\end{equation}
If all the measurement operators are type-$\mathscr{Z}^\perp$, Eq.~(\ref{Eq:IrcohZperp}) summed over $r$ and combined  with Eq.~(\ref{Eq:PAPRdef}) provides an upper bound on mutual information in the form:
\begin{equation}\label{eq:OOK_bound}
\tilde{\sf I} \le \zeta^2 \sum_j p_j |\gamma_j|^2 \log{\cal P} \le \bar{n} \log{\cal P},
\end{equation}
where in the second step we have used Eq.~(\ref{Eq:nbar=}). The above inequality implies that when photon counting detection is used, the PIE is upper bounded by the logarithm of the peak-to-average power ratio ${\cal P}$. It is easy to verify that the maximum value $\log{\cal P}$ can be achieved using a two-symbol ensemble consisting of a coherent state $\ket{\sqrt{{\cal P}\bar{n}}}$ sent with a probability $1/{\cal P}$ and the vacuum state $\ket{0}$ sent with the probability $1-1/{\cal P}$, detected by Geiger-mode photon counting which discriminates between zero photons and at least one photon. Such a strategy is customarily used in deep-space optical communication, where photon information efficiency is of primary importance \cite{Waseda2011, Moision2014}.
The practical implementation is based on the pulse position modulation (PPM) format which uses frames composed of an an integer number of ${\cal P}$ temporal bins. Each frame contains one pulse with the mean photon number ${\cal P}\bar{n}$ located in one of the bins that are otherwise left empty.

Note that Eq.~(\ref{Eq:IrcohZperp}) implies that for PSK alphabets, when all $\gamma_j$ are equal in their magnitude, the contribution to mutual information from type-$\mathscr{Z}^\perp$ operators is zero, $\tilde{\sf I}_r=0$ when $r\in \mathscr{Z}^\perp$. This is consistent with the simple fact that direct photon counting cannot reveal information encoded in the phase of the electromagnetic field. Consequently, a general detection strategy combining type-$\mathscr{Z}$ and type-$\mathscr{Z}^\perp$ operators for a PSK ensemble has the PIE upper bounded by 2 nats per photon in the limit $\bar{n} \rightarrow 0$.

\subsection{Individual measurements on BPSK symbols}

Let us now discuss the case of individual measurements on the binary phase shift keying (BPSK) alphabet of two coherent states $\ket{\zeta}$ and $\ket{-\zeta}$ with opposite phases and the same mean photon number $\bar{n}= \zeta^2$. The maximum mutual information attainable in this scenario can be written in a closed analytical form for a finite $\bar{n}$. We will see that the BPSK alphabet is sufficient to saturate the 2 nats per photon bound on the PIE in the low-cost limit. The exact expression for the mutual information attainable for a given $\bar{n}$ will be used in Sec.~\ref{sec:collective} as a benchmark when discussing superadditive measurements on multiple BPSK symbols.

\begin{figure}
	\centering
	\includegraphics[width=.7\linewidth]{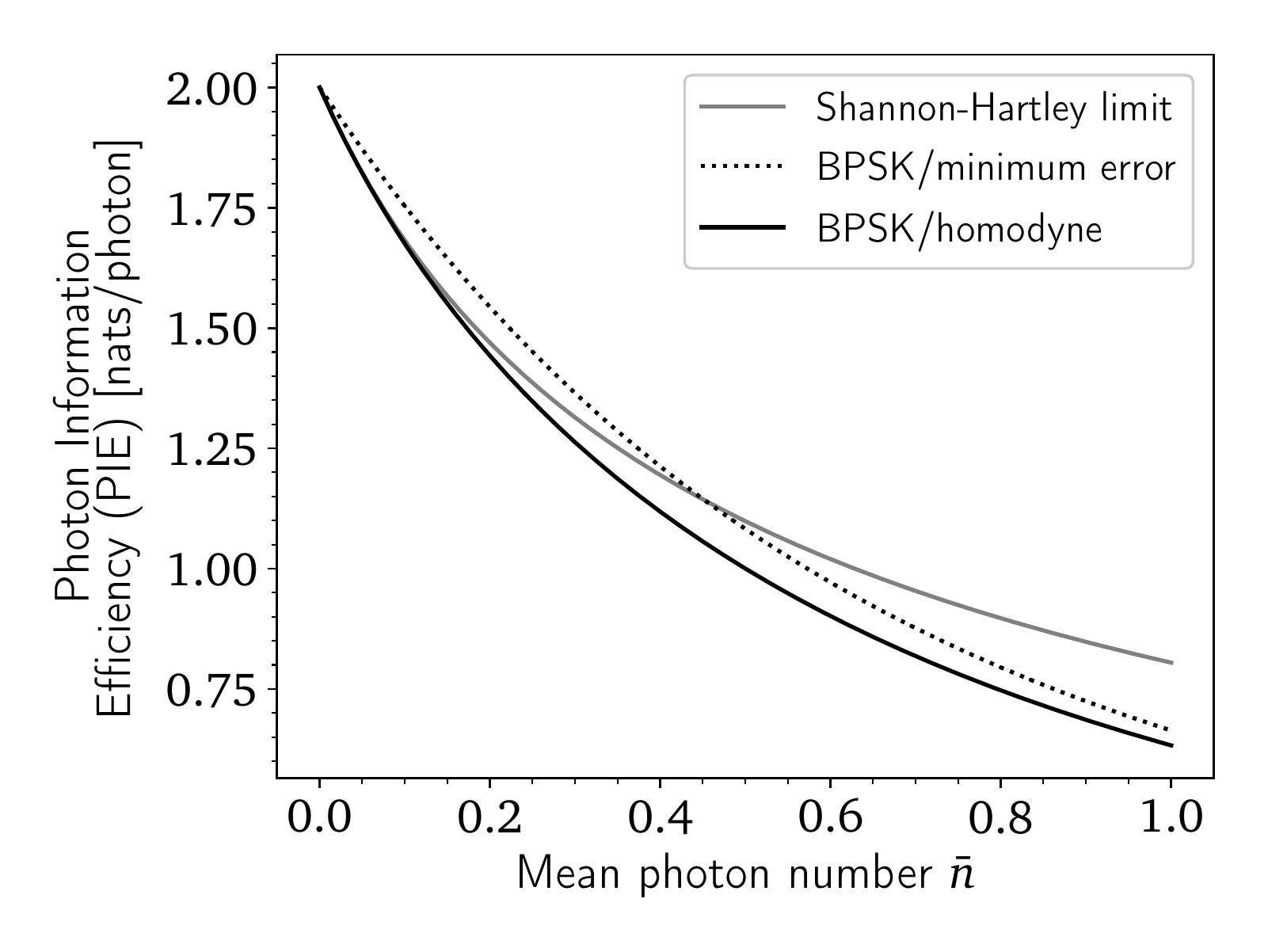}
	\caption{The PIE for the BPSK alphabet detected using the minimum error measurement ${\sf I}_{\textrm{Hel}}/\bar{n}$ (dotted line) compared to the Shannon-Hartley limit ${\sf I}_{\textrm{SH}}/\bar{n}$ (grey solid line), which assumes a Gaussian ensemble of coherent states. The black solid line depicts the case of the homodyne-detected BPSK alphabet ${\sf I}_{\textrm{BPSK/hom}}/\bar{n}$ calculated in Appendix~\ref{Sec:HomodyneScheme}.}
\label{Fig:IDol_SH_Hom}
\end{figure}

It has been shown \cite{OsakiHirotaJMO1998,Tomassoni2008} that when a binary ensemble of equiprobable pure states is used for classical information transmission and only individual detection is permitted, mutual information is maximized by the minimum-error measurement described by Helstrom \cite{Helstrom1976}. The symmetric probability of a wrong identification of the input BPSK state can be obtained from the Helstrom bound
\begin{equation}
\varepsilon_{\text{Hel}} = \frac{1}{2} ( 1- \sqrt{1- |\braket{-\zeta}{\zeta}|^2}) =
\frac{1}{2} ( 1- \sqrt{1- e^{-4\bar{n}}}),
\end{equation}
which yields mutual information expressed in nats:
\begin{equation}
\label{Eq:IHel}
{\sf I}_{\textrm{Hel}} = \log 2-{\sf H}(\varepsilon_{\text{Hel}}).
\end{equation}
Here
\begin{equation}
\label{Eq:BinEntinNats}
{\sf H}(x) = - x \log x - (1-x) \log (1-x)
\end{equation}
is the entropy of a binary random variable specified in nats. The resulting exact expression for the PIE attainable with individual measurements on the BPSK alphabet, equal to ${\sf I}_{\textrm{Hel}}/\bar{n}$, is depicted in Fig.~\ref{Fig:IDol_SH_Hom} as a function of $\bar{n}$. In the limit $\bar{n} \rightarrow 0$ the PIE reaches 2~nats. It is worth noting that for $\bar{n} \lesssim 0.4$ the obtained result exceeds the PIE
${\sf I}_{\textrm{SH}}/\bar{n}$ obtained from the Shannon-Hartley expression given by Eq.~(\ref{Eq:ShannonHartley}), also plotted in Fig.~\ref{Fig:IDol_SH_Hom}. This highlights the specificity of the Shannon-Hartley limit, used as a canonical reference in optical communication, to the quadrature measurement assumed in its derivation. The lower value of ${\sf I}_{\textrm{SH}}/\bar{n}$ compared to ${\sf I}_{\textrm{Hel}}/\bar{n}$ for $\bar{n} \ll 1$ can be verified by inspecting
the second-to-leading terms in the asymptotic expansions of the respective expressions given in Eqs.~(\ref{Eq:ShannonHartley}) and (\ref{Eq:IHel}):
\begin{equation}
{\sf I}_{\textrm{SH}} \approx  2 \bar n - 4 \bar{n}^2, \qquad {\sf I}_{\textrm{Hel}}\approx 2 \bar{n} - \frac{8}{3} \bar{n}^2 .
\label{Eq:IindExpansion}
\end{equation}
As also shown in Fig.~\ref{Fig:IDol_SH_Hom}, the PIE of 2~nats can be attained in the limit $\bar{n} \rightarrow 0$ with a homodyne measurement on the BPSK alphabet, provided that the information is retrieved from the measured continuous quadrature value. For completeness, the mathematical description of this scenario is presented in Appendix~\ref{Sec:HomodyneScheme}. Of course, in this case the PIE does exceed ${\sf I}_{\textrm{SH}}/\bar{n}$ for $\bar{n} >0$.

The physical realization of the minimum error measurement for the BPSK alphabet is known to be the Dolinar receiver based on photon counting of the displaced input signal with a fast feedback loop \cite{Dolinar1973}. Unfortunately, due to its complexity, realizations of the Dolinar receiver have so far been confined to the proof-of-principle stage \cite{Cook2007}. In  search for technically less challenging implementations, alternative nearly-optimal measurements have been investigated, including a feed-forward scheme based on signal splitting and optical delay lines \cite{Sych2016}. A considerable amount of work has been devoted also to displacement receivers without feed-back or feed-forward mechanisms \cite{Wittmann2008,Takeoka2008,Wittmann2010,Tsujino2010,Tsujino2011,DiMario2018,DiMario2019} which show a pretty good performance. Additionally, a non-demolition measurement based on an atom-light interaction that comes close to the Helstrom bound has been recently proposed \cite{Han2018}. Furthermore, derivatives of the Dolinar receiver have been demonstrated for alphabets consisting of more than two coherent states like for example commonly considered quaternary phase shift keying or higher PSK formats \cite{Becerra2011,Izumi2012,Becerra2013,Becerra2015,Ferdinand2017,DiMario2018b,Izumi2020}.
In addition to the minimum-error criterion, one can consider other strategies to distinguish between non-orthogonal quantum states, such as unambiguous state discrimination \cite{Chefles2000,Kilin2001,Barnett2009}. Time-resolved photon counting of the PPM format mentioned in Sec.~\ref{Sec:PIEBounds} can be viewed as a realization of the unambiguous strategy for quantum states of light prepared within PPM frames: registering a photocount in a given time bin identifies unambiguously the location of the pulse, i.e.\ the input state, whereas absence of any photocount over the entire frame yields an inconclusive result.

\section{Collective measurements on BPSK words}
\label{sec:collective}

We will now use the upper bound on the mutual information expanded in the cost parameter as derived in Sec.~\ref{sec:expansion} to identify collective measurement schemes on multiple BPSK symbols that beat the performance of individual detection. While the designs for collective schemes will be motivated by the analysis of the mutual information in the asymptotic limit of the vanishing cost, the actual figure for the mutual information, or equivalently the photon information efficiency, attainable using the designed measurement will be calculated exactly for a given finite $\bar{n}$ and compared against the individual detection benchmark given in Eq.~(\ref{Eq:IHel}).

When collective measurements can be performed on a sequence of individual quantum systems, for example a train of single-mode light pulses, one needs to consider an input ensemble in the form of {\em words} composed from the alphabet of elementary symbols, such as coherent states introduced in Eq.~(\ref{Eq:CoherentStateAlphabet}). The input probabilities are then defined for entire words. In the remainder of the paper, we will restrict our attention to words constructed from the BPSK alphabet of coherent states $\ket{\zeta}$ and $\ket{-\zeta}$. Words of length $M$ can be then conveniently labelled with $M$-bit strings ${\bf j}=j_1 j_2 \ldots j_M$. The input states are pure, $\hat{\varrho}_{\textbf{j}} = \proj{\psi_{\textbf{j}}}$, and have the form of $M$-mode coherent states,
\begin{equation}\label{Eq:Psijseqcoh}
\ket{\psi_{\textbf{j}}} = \ket{(-1)^{j_1} \zeta}\ket{(-1)^{j_2} \zeta} \cdots \ket{(-1)^{j_M} \zeta}.
\end{equation}
The label ${\textbf{j}}$ needs now to be used as an ensemble index in lieu of $j$ when applying general results derived in Sec.~\ref{sec:expansion}. For a fair comparison with the individual measurement scenario, the cost should be quantified with the mean photon number {\em per elementary symbol}, given in the present case by $\bar{n} = \zeta^2$. The entire optical energy of a BPSK word of length $M$ is $M\zeta^2 = M\bar{n}$.

\begin{figure}
	\centering
	\includegraphics[width=.7\linewidth]{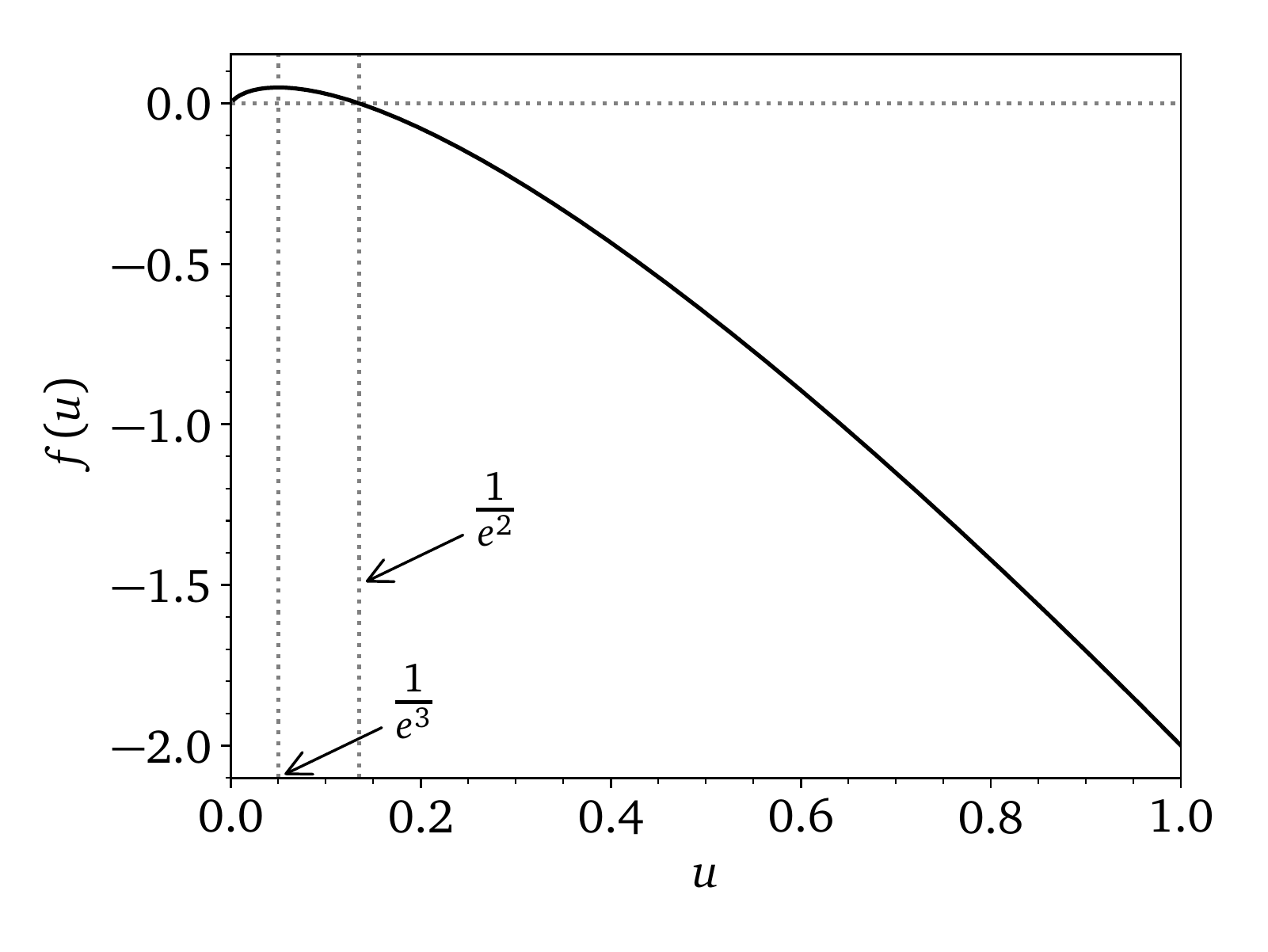}
\caption{The function $f(u)$ defined in Eq.~(\ref{Eq:fdef}). The function is concave and has a single maximum at $u^* = 1/e^3$. For $u \ge 1/e^2$ the function becomes negative.}
\label{Fig:F}
\end{figure}

Our  goal will be to identify the probability distribution $p_{\textbf{j}}$ for input words $\ket{\psi_{\textbf{j}}}$ and the few-symbol collective measurement strategy that maximizes mutual information in the low-cost limit. Let us first specialize Eq.~(\ref{eq:mutinf_expans}) to the scenario considered here. It will be convenient to denote a normalized superposition state of one photon in $M$ modes as:
\begin{align}\label{eq:singlePhotonStates}
\ket{1_{\textbf{j}}} & = \frac{1}{\sqrt{M}} [(-1)^{j_1} \ket{1}\ket{0}\ldots \ket{0}
 + (-1)^{j_2}\ket{0}\ket{1}\ldots \ket{0} + \cdots (-1)^{j_M}\ket{0}\ket{0}\ldots \ket{1}]
\end{align}
and use $\ket{\text{v}} = \ket{0}\ket{0}\ldots \ket{0}$ for the $M$-mode vacuum state. With this notation, the SLD difference  for the state $\hat{\varrho}_{\textbf{j}}$  reads
\begin{equation}
\hat{D}_{\textbf{j}} = 2\sqrt{M}\bigl( \ket{1_{\textbf{j}}}\bra{\text{v}} + \ket{\text{v}}\bra{1_{\textbf{j}}}\bigr)
\end{equation}
and respectively
\begin{equation}
\hat{\varrho}'^{(2)}_{\textbf{j}} = M \proj{1_{\textbf{j}}}.
\end{equation}
After inserting these expressions into Eq.~(\ref{eq:mutinf_expans}) it is easy to notice that pairs of terms corresponding to $\textbf{j}$ and its bitwise negation $\neg\textbf{j}$ can be combined together because the corresponding one-photon states $\ket{1_\textbf{j}}$ and $\ket{1_{\neg\textbf{j}}}$ differ only by an irrelevant global minus sign. We will select one of the $\textbf{j}$'s as a representative for each pair and use
the equivalence class $[\textbf{j}]$ as an index for the two-fold reduced sum over the input ensemble. The result reads
\begin{align}\label{eq:mut1}
\tilde{\sf I} \le 2 M \zeta^2 + M \zeta^2 {\sum_{r \in \mathscr{Z}^\perp}}{\sum_{[{\textbf j}]}} p_{[{\textbf j}]} \bra{1_{[{\textbf j}]}}\hat{Q}_r \ket{1_{[{\textbf j}]}}
\left(\log
\frac{\bra{1_{[{\textbf j}]}}\hat{Q}_r \ket{1_{[{\textbf j}]}}}%
{\sum_{[{\textbf l}]} p_{[{\textbf l}]} \bra{1_{[{\textbf l}]}}\hat{Q}_r \ket{1_{[{\textbf l}]}}}
-2\right),
\end{align}
where $p_{[{\textbf j}]}= p_{\textbf j} + p_{\neg{\textbf j}}$ is the total probability of using either word from a given pair. The above expression can be equivalently written as
\begin{align}\label{eq:calc}
\tilde{\sf I} &\le 2 M \zeta^2 + M \zeta^2 {\sum_{r \in \mathscr{Z}^\perp}} \sum_{[{\textbf j}]} f(p_{[{\textbf j}]})
\bra{1_{[{\textbf j}]}}\hat{Q}_r  \ket{1_{[{\textbf j}]}}\\
&+ M\zeta^2
\sum_{r \in \mathscr{Z}^\perp}\sum_{[{\textbf j}]}p_{[{\textbf j}]} \bra{1_{[{\textbf j}]}}\hat{Q}_r \ket{1_{[{\textbf j}]}}\log
\frac{p_{[{\textbf j}]}\bra{1_{[{\textbf j}]}}\hat{Q}_r \ket{1_{[{\textbf j}]}}}{\sum_{{[{\textbf l}]}} p_{{[{\textbf l}]}} \bra{1_{{[{\textbf l}]}}}\hat{Q}_r \ket{1_{{[{\textbf l}]}}}},
\end{align}
where
\begin{equation}
\label{Eq:fdef}
f(u) = u \log\frac{1}{u} - 2u.
\end{equation}
Since for any ${\textbf j}$ and $r$ the argument of the logarithm in Eq.~(\ref{eq:calc}) does not exceeds one, the second line of that expression containing logarithms is nonpositive therefore by neglecting it one arrives at a weaker upper bound of the form:
\begin{equation}\label{eq:upbound}
\tilde{\sf I} \le 2 M \zeta^2 + M \zeta^2 \sum_{[{\textbf j}]} f(p_{[{\textbf j}]})
\bra{1_{[{\textbf j}]}}  \left( {\sum_{r\in \mathscr{Z}^\perp}} \hat{Q}_r  \right) \ket{1_{[{\textbf j}]}}.
\end{equation}
The function $f(u)$ defined in Eq.~(\ref{Eq:fdef}) will play an important role in further analysis. Its graph is shown in Fig.~\ref{Fig:F}. It is easy to verify by an elementary calculation that on the interval $0\le u \le 1$ the function $f(u)$ is concave, has a single maximum at $u^\ast=1/e^3\approx 0.0498$ equal to $f(u^\ast) = u^\ast$ and becomes negative for arguments $u> 1/e^2$.

\subsection{Two-symbol measurements}
\label{Subsec:2copy}

In the case of two-symbol measurements, $M=2$, we have only two non-equivalent one-photon states that can be compactly denoted as
\begin{align}
\ket{+} & = \ket{1_{00}} = - \ket{1_{11}} = \frac{1}{\sqrt{2}}(\ket{1}\ket{0} + \ket{0}\ket{1}), \\
\ket{-} & = \ket{1_{01}} = - \ket{1_{10}} = \frac{1}{\sqrt{2}}(\ket{1}\ket{0} - \ket{0}\ket{1}).
\end{align}
We will also use $+$ and $-$ symbols to label the two respective equivalence classes $\mathord{+} \equiv [00]=[11]$ and $\mathord{+} \equiv [01]=[10]$ of the input words with corresponding probabilities $p_+ = p_{00} + p_{11}$ and
$p_- = p_{10} + p_{01}$. Eq.~(\ref{eq:upbound}) can be then simplified to the form
\begin{equation}
\label{Eq:IMr-Mr+}
\tilde{\sf I}  \le 2\bar{n} [ 2+ Q^+ f(p_+) + Q^- f(p_-)],
\end{equation}
where we have denoted
\begin{equation}
Q^+ = \sum_{r \in \mathscr{Z}^\perp} \bra{+} \hat{Q}_r \ket{+}, \quad Q^- = \sum_{r \in \mathscr{Z}^\perp} \bra{-} \hat{Q}_r \ket{-}.
\end{equation}
Because $Q^\pm$ are the overall probabilities of obtaining an $r\in \mathscr{Z}^\perp$ outcome on states $\ket{\pm}$, they are constrained by $0 \le Q^\pm \le 1$.
The goal now is to maximize the right hand side of Eq.~(\ref{Eq:IMr-Mr+}). Given the properties of the function $f(u)$ discussed earlier and taking into account that $p_++p_-=1$, the right hand side of Eq.~(\ref{Eq:IMr-Mr+}) is optimized by selecting $p_- = u^\ast=1/e^3$, $Q^-=1$, and $Q^+=0$. An equivalent solution has exchanged labels $+$ and $-$. This finally yields:
\begin{equation}
\label{Eq:Itwosymbolfinal}
\tilde{\sf I} \le 2 \bar{n} [2+ f(u^\ast)] = 2 \bar{n} (2+ 1/e^3).
\end{equation}
Given that the mean photon number in an input word is $2\bar{n}$, the asymptotic value of the PIE is upper bounded by $\tilde{\sf I}/(2\bar{n}) \le 2+1/e^3$, which presents a relative increase of $1/(2e^3) \approx 2.49\%$ compared to the individual measurement scenario. Note that the value derived in Eq.~(\ref{Eq:Itwosymbolfinal}) maximizes also the full expression given in Eq.~(\ref{eq:calc}), as for the chosen $p_{\pm}$ and $Q_{\pm}$ the second logarithmic term in Eq.~(\ref{eq:calc}) vanishes.

The optimum identified above can be translated into an optical measurement scheme that asymptotically saturates the  bound derived in Eq.~(\ref{Eq:Itwosymbolfinal}). The condition $Q^-=1$ implies that input states $\ket{\psi_{01}}$ and $\ket{\psi_{10}}$ containing the $\ket{-}$ component should be detected by type-$\mathscr{Z}^\perp$ measurement operators, which as discussed in Sec.~\ref{Sec:Individual} correspond in the optical domain to single photon detection. Since photon detection is phase-insensitive, the optimum $p_- = p_{01} + p_{10} $ specifies only the overall probability of using the two input states $\ket{\psi_{01}}$ and $\ket{\psi_{10}}$. For concreteness, we will take only $\ket{\psi_{01}}$. On the other hand, the requirement $Q^+=0$ means that input states $\ket{\psi_{00}}$ and $\ket{\psi_{11}}$ containing the one-photon component $\ket{+}$ are measured only by type-$\mathscr{Z}$ operators, which corresponds to phase-sensitive detection. We have seen in Sec.~\ref{Sec:Individual} that such a measurement can be realized by the Dolinar receiver or homodyning. In order to maximize the contribution to mutual information, one needs to use the states $\ket{\psi_{00}}$ and $\ket{\psi_{11}}$ with the same input probabilities $p_{00} = p_{11}$.

\begin{figure}
	\centering
	\includegraphics[width=.7\linewidth]{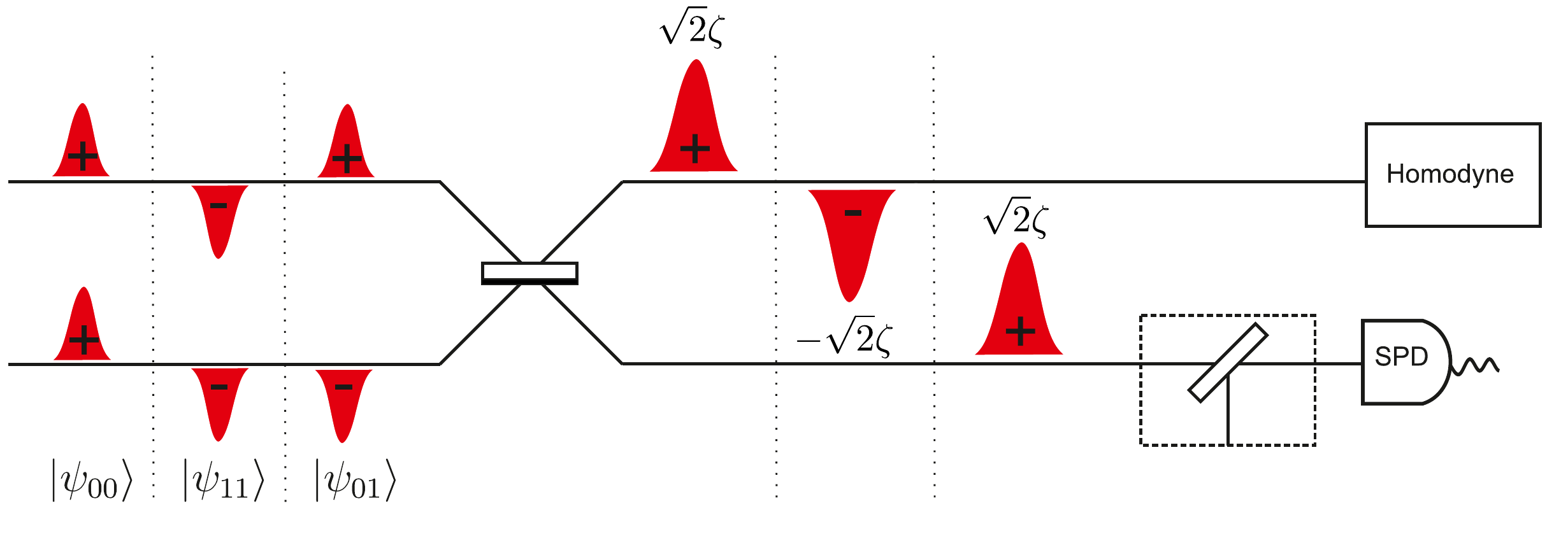}
	\caption{A communication scheme providing superadditive advantage for two-symbol BPSK words. Input states $\ket{\psi_{00}},\,\ket{\psi_{11}},\,\ket{\psi_{01}}$ are interfered on a 50:50 beamsplitter resulting in the transformation of Eq.~(\ref{Eq:2symbolTransform}). The first output mode undergoes homodyne detection whereas the second one is measured by a single photon detector (SPD). The optional displacement operation preceding the SPD can slightly enhance mutual information.}
	\label{Fig:2copy_detect}
\end{figure}

\begin{figure}
	\centering
	\includegraphics[width=.5\linewidth]{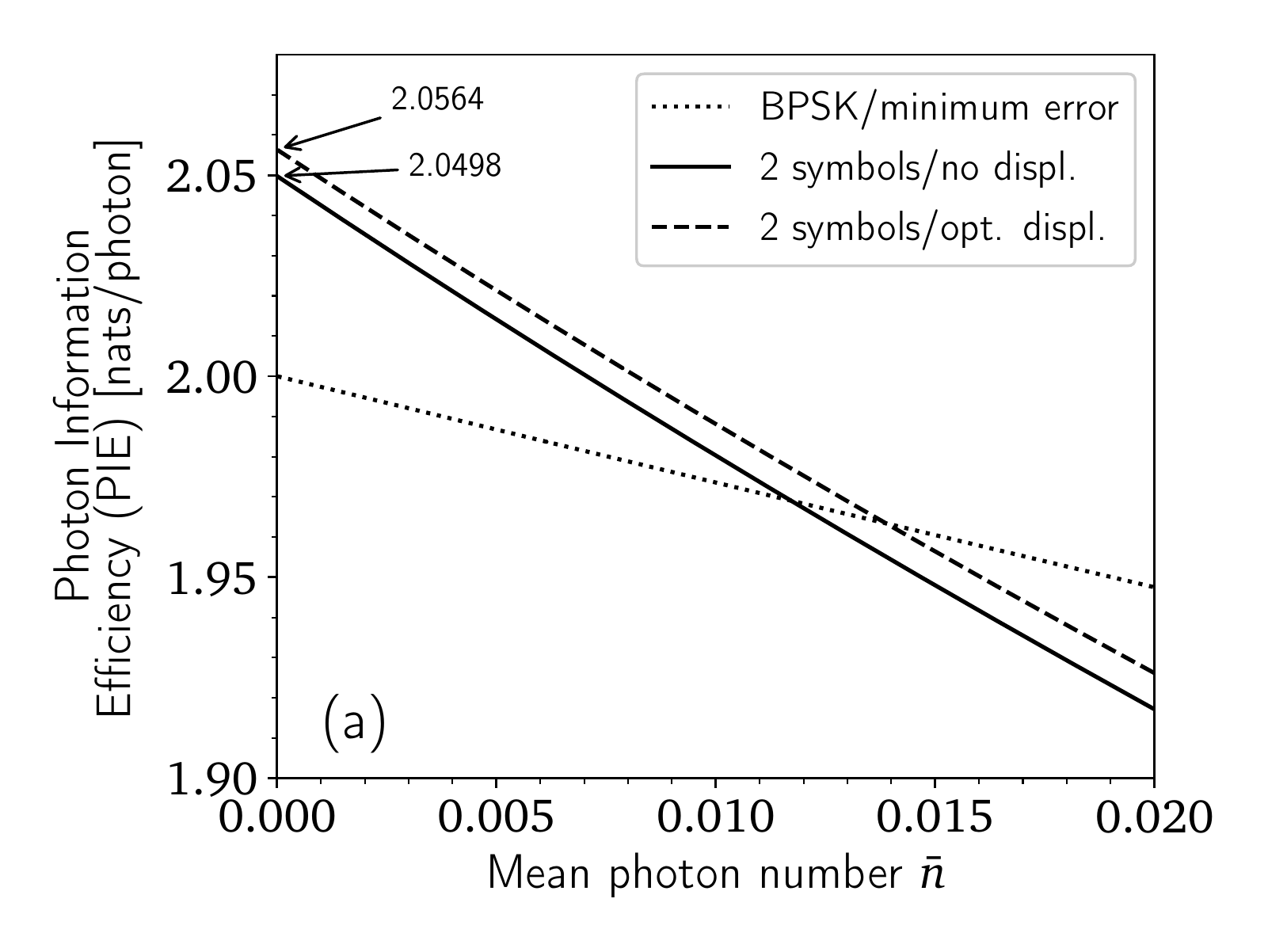}
\includegraphics[width=.49\linewidth]{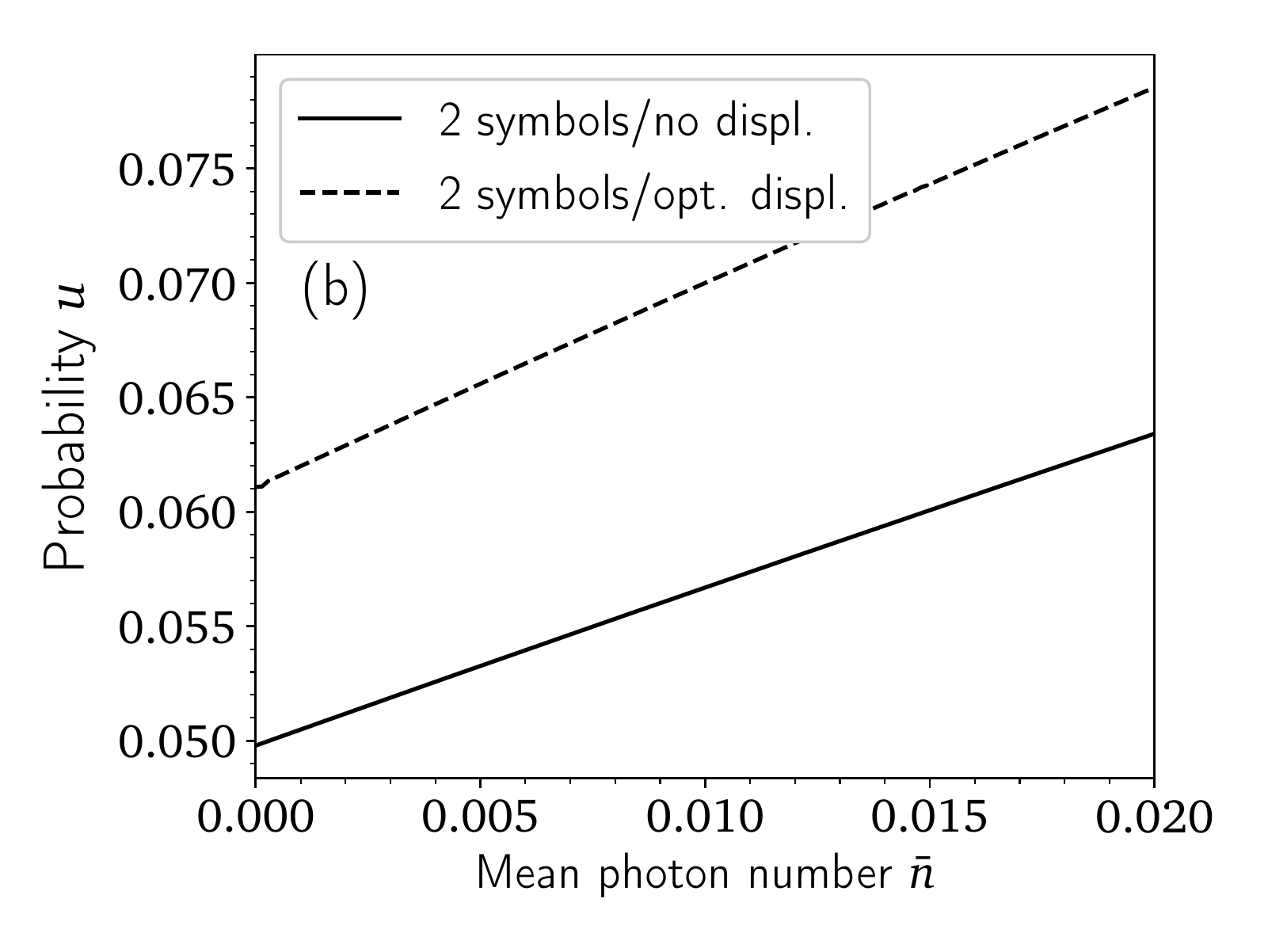}
	\caption{{\bf(a)} The PIE defined as ${\sf I_2}/(2\bar{n})$ for the two-symbol measurement shown schematically in  Fig.~\ref{Fig:2copy_detect} (solid line, without displacement preceding SPD; dashed line, with displacement) compared to the optimal individual measurement (dotted line) as a function of the mean photon number $\bar n$. Without a displacement operation  the PIE tends to 2.0498 (an advantage of 2.49\%) as derived in Eq.~(\ref{Eq:Itwosymbolfinal}). With displacement the PIE reaches at 2.0564 (an advantage of 2.82\%).  {\bf(b)} The probability $u$ of sending the state $\ket{\psi_{01}}$ in the scheme without (solid line) or with (dashed line) displacement as a function of the mean photon number $\bar n$.}
	\label{Fig:2CopyOptimization}
\end{figure}

A straightforward way to subject the input states to the required type of detection is to interfere the two component coherent states in each two-symbol word on a balanced 50/50 beam splitter, as shown in Fig.~\ref{Fig:2copy_detect}. This results in the transformation
\begin{align}\label{Eq:2symbolTransform}
\ket{\psi_{00}} & \rightarrow \ket{\sqrt{2}\zeta}\ket{0}, \nonumber \\
\ket{\psi_{11}} & \rightarrow \ket{-\sqrt{2}\zeta}\ket{0}, \nonumber \\
\ket{\psi_{01}} & \rightarrow \ket{0}\ket{\sqrt{2}\zeta} ,
\end{align}
where the two kets on the right hand side refer to the two output ports of the beam splitter. Note that when a Dolinar receiver is used as a phase-sensitive measurement, this reproduces the two-symbol measurement strategy described by Guha \cite{Guha2011}. In the following, we consider homodyne detection as a more standard phase-sensitive measurement technique. For a finite $\bar{n}=\zeta^2$ we have optimized numerically the PIE ${\sf I_2}/(2\bar{n})$ given by the ratio  of the actual mutual information ${\sf I_2}$ calculated using Eq.~(\ref{Eq:I2App}) in Appendix~\ref{Sec:HomodyneScheme} and the mean photon number over input probabilities parameterized with $p_{01}+p_{10} = u$, $p_{00} = p_{11} = (1-u)/2$ in a scenario when a combination of homodyne detection and single photon detection (SPD) is implemented on the two output beam splitter ports. The results are shown in Fig.~\ref{Fig:2CopyOptimization}. It is seen that in the limit $\bar{n}\rightarrow 0$ the PIE indeed approaches the value derived in Eq.~(\ref{Eq:Itwosymbolfinal}) which is also shown analytically in Appendix~\ref{Sec:HomodyneScheme}. Superadditivity of accessible information for the BPSK alphabet, when the amount of transmitted information exceeds that attainable with individual measurements, occurs for $\bar{n} < 0.0117$.

\begin{figure}
	\centering
	\includegraphics[width=.7\linewidth]{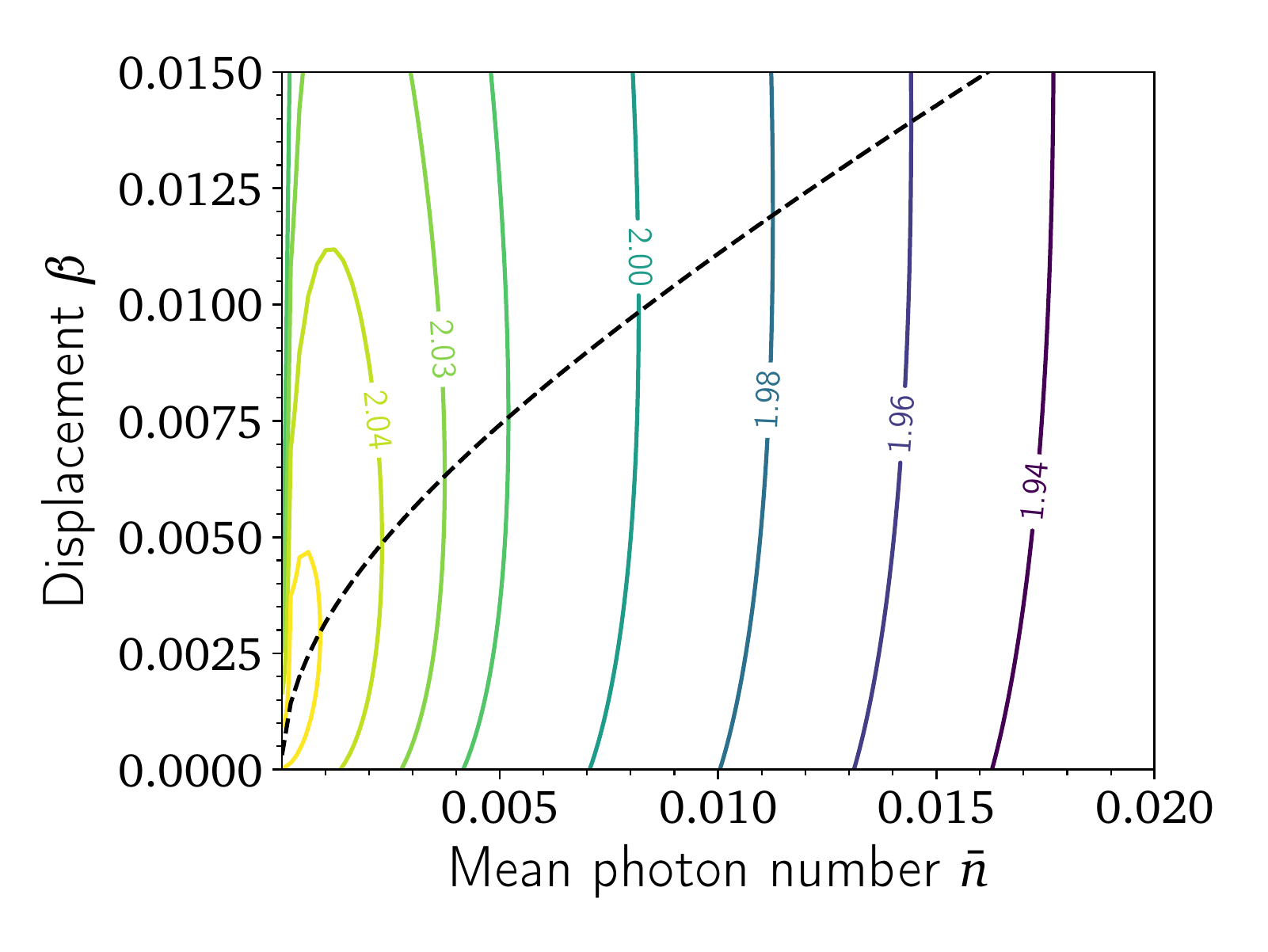}
	\caption{A contour plot of the PIE ${\sf I}/(2\bar{n})$ for the two-symbol measurement depicted in Fig.~\ref{Fig:2copy_detect} as a function of the mean photon number $\bar n$ and the displacement $\beta$. The dashed line indicates the optimal displacement value for a given mean photon number $\bar n$.}
	\label{Fig:2CopyContour}
\end{figure}

The PIE attainable for two-symbol measurements can be further improved by introducing a displacement operation before the single photon detector (SPD), shown within a dashed box in Fig.~\ref{Fig:2copy_detect}. This operation adds a coherent amplitude $\beta$ to the field detected by the SPD. Such an arrangement  is known as the generalized Kennedy receiver \cite{Wittmann2010, Tsujino2011, DiMario2018, DiMario2019}. As seen in Fig.~\ref{Fig:2CopyOptimization}, optimizing $\beta$ for a given $\bar{n}$ yields a slightly higher value of PIE, approaching $2.0564$ in the limit $\bar{n}\rightarrow 0$, which corresponds to $2.82\%$ enhancement relative to individual measurements. Noticeably, the displacement value $\beta$ tends to zero with diminishing $\bar{n}$, as seen in Fig.~\ref{Fig:2CopyContour}. This result indicates that the asymptotic analysis of mutual information in the low-cost limit carried out in Sec.~\ref{sec:expansion} may yield a different bound if explicit dependence of the measurement operators on the cost parameter is allowed. A similar effect occurs also in the analysis of superadditivity of accessible information in the case of a collective measurement on qubit pairs \cite{Buck1999}.

\subsection{Three-symbol measurements}

In the case of three-symbol BPSK words, the one-photon sector relevant to the evaluation of mutual information in the low-cost limit is spanned by four states:
\begin{align}
\ket{\mathord{+}\mathord{+}} & = \ket{1_{000}} = - \ket{1_{111}} = \frac{1}{\sqrt{3}}(\ket{100}+\ket{010}+\ket{001}), \nonumber \\
\ket{\mathord{+}\mathord{-}} & = \ket{1_{001}} = - \ket{1_{110}} = \frac{1}{\sqrt{3}}(\ket{100}+\ket{010}-\ket{001}), \nonumber \\
\ket{\mathord{-}\mathord{+}} & = \ket{1_{010}} = - \ket{1_{101}}= \frac{1}{\sqrt{3}}(\ket{100}-\ket{010}+\ket{001}), \nonumber \\
\ket{\mathord{-}\mathord{-}} &
= \ket{1_{011}} = - \ket{1_{100}} = \frac{1}{\sqrt{3}}(\ket{100}-\ket{010}-\ket{001}).
\label{eq:3statesSingle}
\end{align}
An essential difference with the two-symbol case is that the four single-photon states listed above are mutually nonorthogonal. This makes it problematic to use Eq.~(\ref{eq:upbound}) for optimization as
the matrix elements $\bra{1_{[\textbf{j}]}} \bigl( \sum_{r\in \mathscr{Z}^\perp} \hat{Q}_r \bigr) \ket{1_{[\textbf{j}]}}$ cannot be set independently for each one photon state $\ket{1_{[\textbf{j}]}}$. Consequently, one needs to revert back to Eq.~(\ref{eq:mut1}) for further analysis.
We will denote the four equivalence classes with indices $\mathord{+}\mathord{+}\equiv[000]=[111]$, $\mathord{+}\mathord{-}\equiv[001]=[110]$, $\mathord{-}\mathord{+} \equiv [010]=[101]$, and $\mathord{-}\mathord{-} \equiv [011]=[100]$.

\begin{figure}
	\centering
	\includegraphics[width=.7\linewidth]{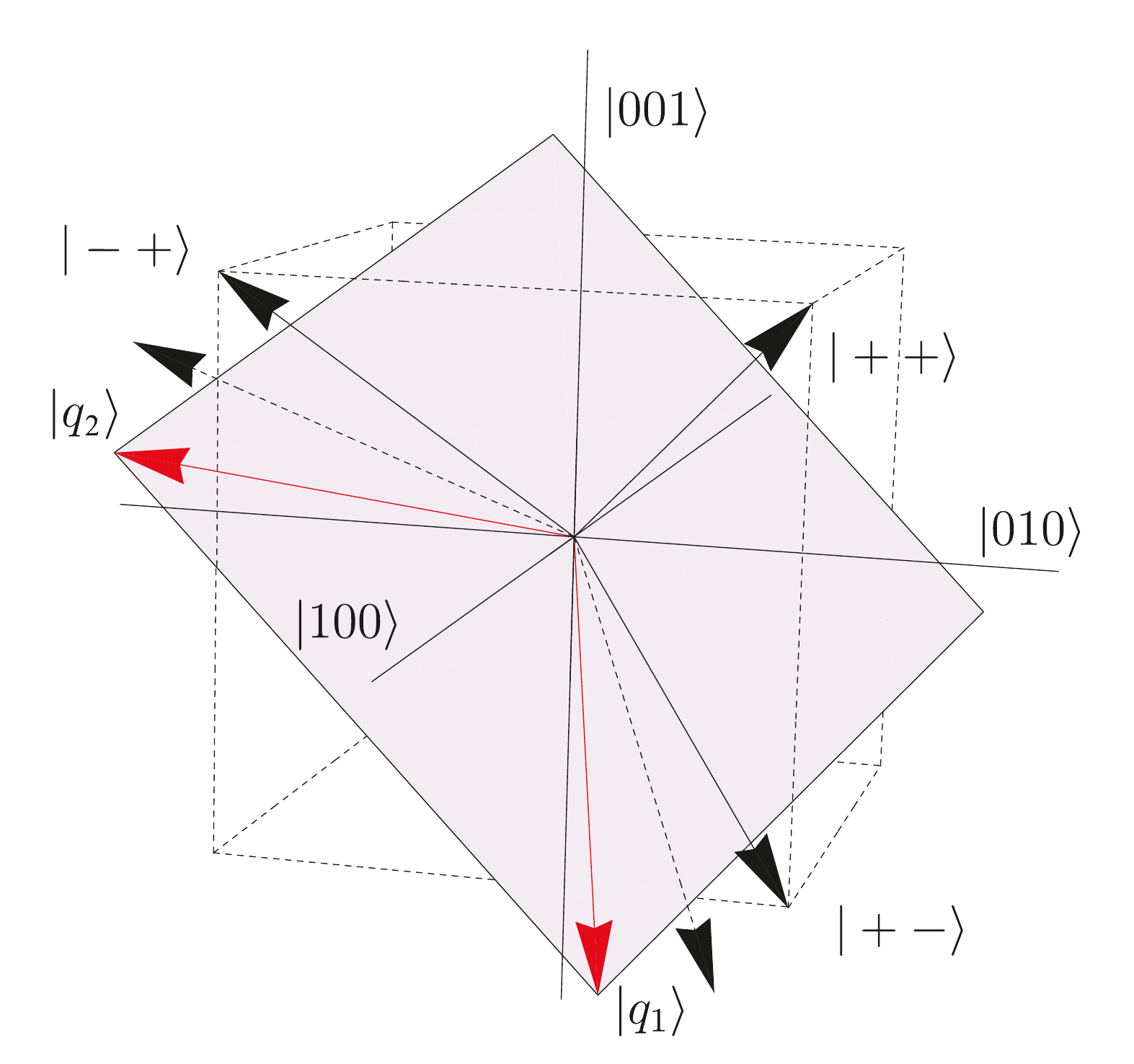}
	\caption{Graphic representation of the optimal measurement attaining superadditive advantage for three-symbol BPSK words. Input states $\ket{\psi_{000}},\,\ket{\psi_{001}},\,\ket{\psi_{010}}$ corresponding to single-photon states $\ket{++},\,\ket{+-},\,\ket{-+}$ represented by black arrows are measured using a projective POVM with one element parallel to $\ket{++}$ and two others being $\ket{q_0}$ and $\ket{q_1}$, indicated by red arrows. The latter part of the POVM is the minimum error measurement between projections of $\ket{+-},\,\ket{-+}$ onto plane orthogonal to $\ket{++}$, indicated by dashed arrows.}
	\label{Fig:3copy_measure}
\end{figure}

We have performed a numerical search for input probabilities $p_{[{\textbf{j}}]}$ and type-$\mathscr{Z}^\perp$ POVM elements of rank-$1$ form in the one-photon sector that maximize the right hand side of Eq.~(\ref{eq:mut1}). Up to a permutation of the modes, the numerical results suggest an optimum characterized by a high degree of symmetry with
\begin{equation}
	p_{++} = 1 - 2 v , \qquad p_{+-} = p_{-+} = v , \qquad p_{--} = 0
	\label{Eq:3copy_prob}
\end{equation}
and two type-$\mathscr{Z}^\perp$ projective operators $\hat{Q}_r = \proj{q_r}$, $r=1,2$ that are given in the orthonormal basis $\ket{100}, \ket{010}, \ket{001}$ by
\begin{equation}
		\ket{q_1} = \begin{pmatrix}\frac{1}{\sqrt{3}} \\ \frac{1}{2} (1-\frac{1}{\sqrt{3}}) \\ - \frac{1}{2} (1+\frac{1}{\sqrt{3}})\end{pmatrix} , \quad
		\ket{q_2} =  \begin{pmatrix} \frac{1}{\sqrt{3}} \\ -\frac{1}{2} (1+\frac{1}{\sqrt{3}}) \\ \frac{1}{2} (1-\frac{1}{\sqrt{3}})\end{pmatrix} .  \label{Eq:3copy_meas}
\end{equation}
It is straightforward to verify that $\braket{q_1}{\mathord{+}\mathord{+}} = \braket{q_2}{\mathord{+}\mathord{+}}= 0$.
The geometry of the measurement is visualized in Fig.~\ref{Fig:3copy_measure} with a three-dimensional diagram using the fact that the relevant state vectors have real components. The four one-photon states specified in Eq.~(\ref{eq:3statesSingle}) are located in vertices of a cube centered at the origin of the coordinate system. The two mutually orthogonal states $\ket{q_1}$ and $\ket{q_2}$ corresponding to projective type-$\mathscr{Z}^\perp$ operators lie in the plane perpendicular to $\ket{\mathord{+}\mathord{+}}$.
This means that the vector $\ket{\mathord{+}\mathord{+}}$ defines a subspace on which
a type-$\mathscr{Z}$ measurement needs to be carried out. The vectors $\ket{q_1}$ and $\ket{q_2}$ can be viewed as an implementation of a minimum-error measurement for inputs $\ket{\mathord{+}\mathord{-}}$ and $\ket{\mathord{-}\mathord{+}}$ projected  onto the plane perpendicular to $\ket{\mathord{+}\mathord{+}}$.

Inserting Eq.~(\ref{Eq:3copy_prob}) and type-$\mathscr{Z}^\perp$ operators specified in Eq.~(\ref{Eq:3copy_meas}) into Eq.~(\ref{eq:calc}) yields
\begin{align}\label{Eq:3copy_bound}
	\tilde{\sf I}_3 \le 6\bar{n}+\frac{6\bar n}{9} \{ 4 v [\sqrt{3} \log(2+\sqrt{3})-4\log 2-v]
+ 8 f(v) \}.
\end{align}
The right hand side reaches maximum at $v^\ast \approx 0.0375$ resulting in a PIE bound
$\tilde{\sf I}_3/(3\bar{n}) \le 2.0679$, which represents relative $3.40\%$ improvement compared to individual measurements.

\begin{figure}
	\centering
	\includegraphics[width=.7\linewidth]{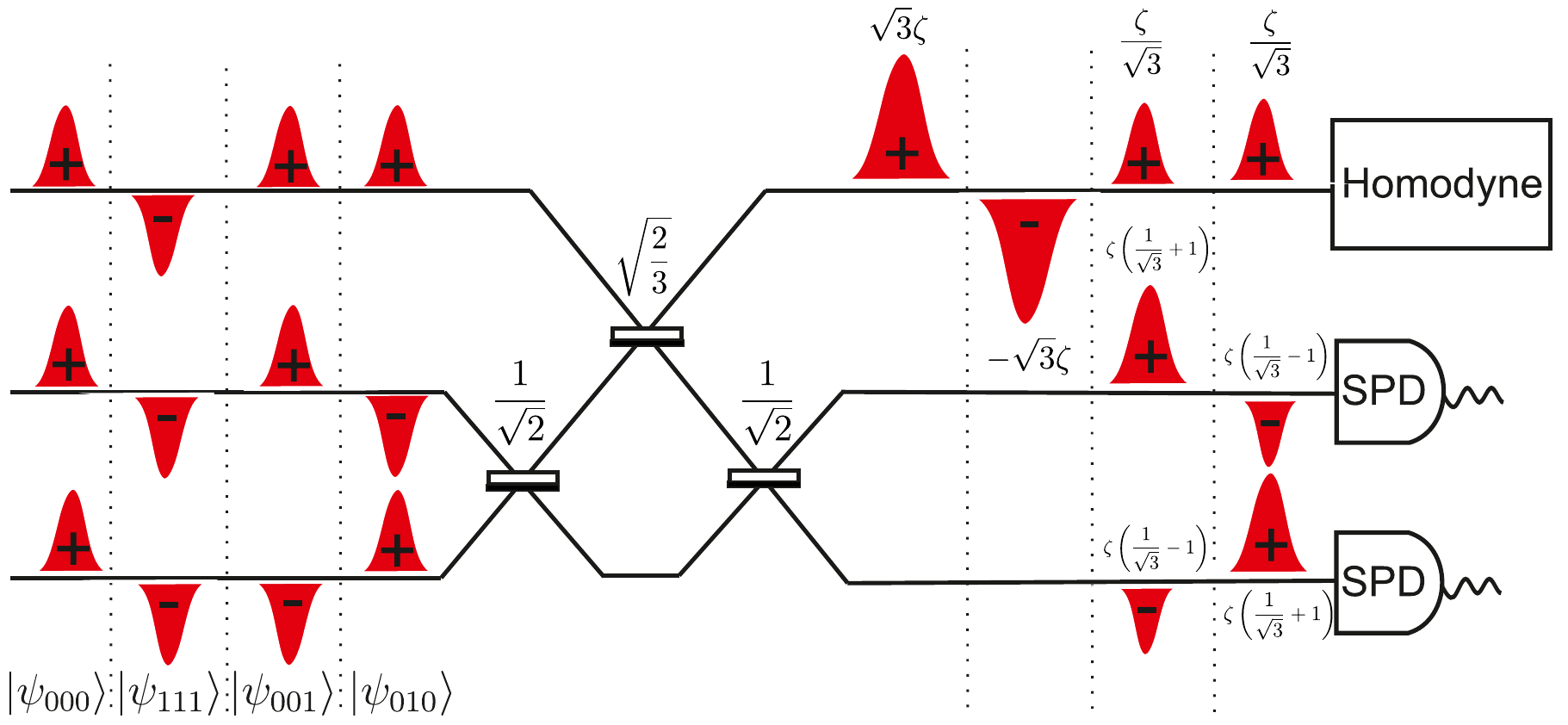}
	\caption{A schematic representation of a communication protocol attaining superadditive advantage for three-symbol BPSK words. Input states $\ket{\psi_{000}},\,\ket{\psi_{111}},\,\ket{\psi_{001}},\,\ket{\psi_{010}}$ are first mixed by a set of beam splitters with respective transmissivities $50\%$ and $66\%$ resulting in respective states. The first output mode undergoes a homodyne detection whereas the second and the third one are measured by SPDs.}
	\label{Fig:3copy_detect}
\end{figure}

\begin{figure}
	\centering
	\includegraphics[width=.5\linewidth]{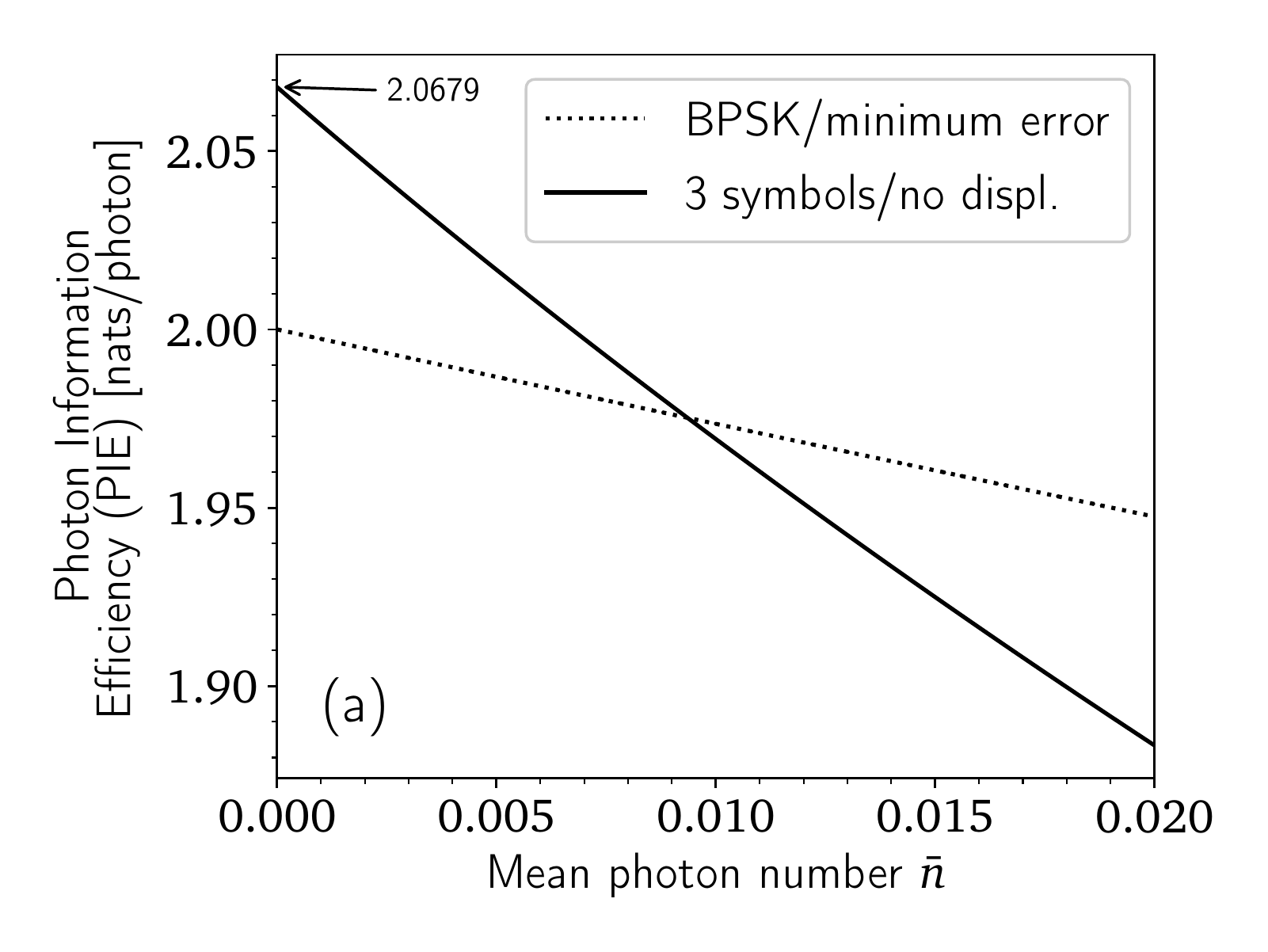}
	\includegraphics[width=.49\linewidth]{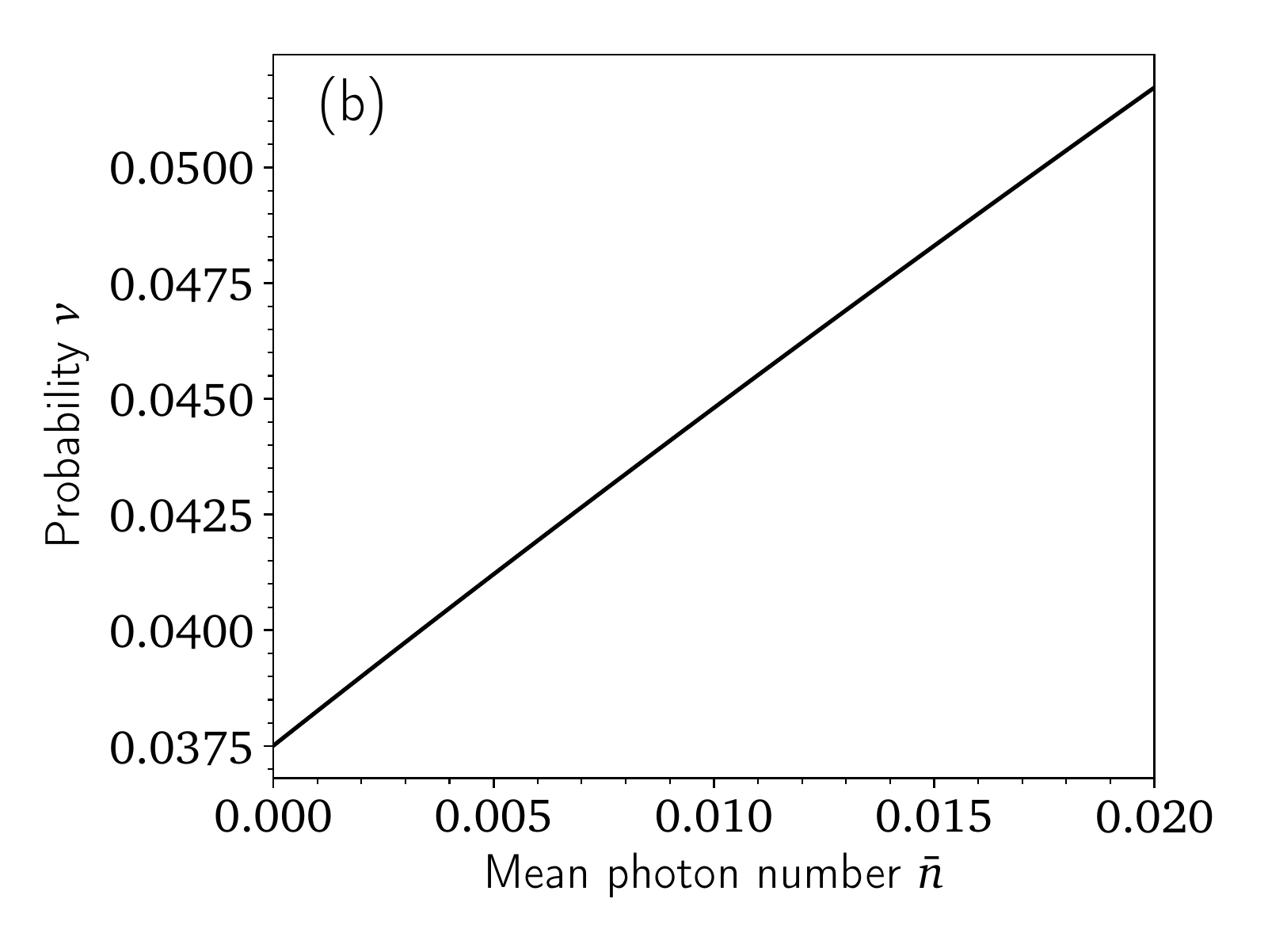}
	\caption{{\bf (a)} The PIE defined as ${\sf I}/(3\bar{n})$ of the three-symbol measurement depicted in Fig.~\ref{Fig:3copy_detect} (solid line) compared to the optimal individual measurement (dotted line) as a function of the mean photon number $\bar n$. {\bf (b)} The probability $v$ of sending either the state $\ket{\psi_{001}}$ or $\ket{\psi_{010}}$ which are detected by a SPD as a function of the mean photon number $\bar n$.}
	\label{Fig:3copyOptimization}
\end{figure}

Analogously to the two-symbol case, the above solution suggests an optical implementation shown in Fig.~\ref{Fig:3copy_detect}. A single three-port linear optical circuit is constructed such that photons prepared in one of three orthogonal superoposition states $\ket{\mathord{+}\mathord{+}}$, $\ket{q_1}$, and $\ket{q_2}$ are routed to different output ports.
Homodyne detection is implemented on the port corresponding to the $\ket{\mathord{+}\mathord{+}}$ input, while two other ports are monitored by SPDs.
The input probabilities are parameterized as $p_{000} = p_{111} = 1/2-v$ and $p_{001}=p_{010}=v$. Maximization over $v$ for a given $\bar{n}=\zeta^2$ yields the  actual photon information efficiency ${\sf I}_3/(3\bar{n})$ shown in Fig.~\ref{Fig:3copyOptimization}(a) with the optimal value of $v$ depicted in Fig.~\ref{Fig:3copyOptimization}(b). It is seen that the PIE and and the probability $v$ attain respectively the $3.40\%$ enhancement and the optimal value implied by Eq.~(\ref{Eq:3copy_bound}). in the asymptotic limit $\bar{n} \rightarrow 0$.

\subsection{Orthogonal words}
Although for BPSK words longer than two symbols full optimization of the asymptotic expression for mutual information requires resorting to numerical means, some interesting observations can be made when the ensemble of input words is {\em a priori} restricted to a subset for which all the one-photon states defined in Eq.~(\ref{eq:singlePhotonStates}) are mutually orthogonal, apart from pairs that differ only by a global minus sign. Such subsets can be systematically constructed using Hadamard matrices, which are real orthogonal matrices with entries $\pm 1$ defined in dimensions that are integer powers of $2$ and conjectured to exist for any dimension that is a multiple of $4$ \cite{Horadam2007}. The orthogonal words are chosen such that
sequences of signs for individual coherent states specified in Eq.~(\ref{Eq:Psijseqcoh}) are given by rows of a Hadamard matrix, up to a global phase flip.
The orthogonality property of Hadamard matrices mean that for such a reduced subset one-photon states $\ket{1_{\textbf j}}$ and $\ket{1_{{\textbf j}'}}$ defined in Eq.~(\ref{eq:singlePhotonStates}) are mutually orthogonal when ${\textbf j}$ and ${{\textbf j}'}$ correspond to different rows of a Hadamard matrix.

The orthogonality of the states $\ket{1_{[{\textbf j}]}}$ implies that the value of each of the matrix elements $Q^{[{\textbf j}]} = \bra{1_{[{\textbf j}]}}  \left( {\sum_{r\in \mathscr{Z}^\perp}} \hat{Q}_r  \right) \ket{1_{[{\textbf j}]}}$ in Eq.~(\ref{eq:upbound}) can be set independently within physical constraints $0 \le Q^{[{\textbf j}]} \le 1$. The obvious way to maximize Eq.~(\ref{eq:upbound}) is then to take $Q^{[{\textbf j}]} =1$ if $f(p_{[{\textbf j}]})>0$ and $Q^{[{\textbf j}]} =0$ otherwise. Recalling that $f(t)$ is positive for arguments $0< t < 1/e^2$,  this reduces Eq.~(\ref{eq:upbound})
to
\begin{equation}
\tilde{\sf I} \le 2 M \zeta^2 + M \zeta^2
\sum_{[{\textbf j}]; \;  0 < p_{[{\textbf j}]} < 1/e^2}
f(p_{[{\textbf j}]}).
\label{Eq:IHadamard}
\end{equation}
Importantly, by taking measurement operators $\hat{Q}_r$ in the form of projections onto one-photon states $\proj{1_{[{\textbf j}]}}$, the second line of Eq.~(\ref{eq:calc}) can be made identically equal to zero, which implies that the bound (\ref{Eq:IHadamard}) is equivalent to Eq.~(\ref{eq:mut1}).

Suppose now that exactly $M'$ indices $[{\textbf j}]$ satisfy the condition $0 < p_{[{\textbf j}]} < 1/e^2$. The concavity of $f(t)$ on the interval $0< t < 1/e^2$ implies that
\begin{equation}
\sum_{[{\textbf j}]; \;  0 < p_{[{\textbf j}]} < 1/e^2}
f(p_{[{\textbf j}]}) \le M' f\left( \frac{1}{M'} \sum_{[{\textbf j}]; \;  0 < p_{[{\textbf j}]} < 1/e^2}
p_{[{\textbf j}]}\right).
\end{equation}
This means that it is always beneficial to make all the probabilities $p_{[{\textbf j}]}$ from the interval $0 < p_{[{\textbf j}]} < 1/e^2$ identical. Consequently, the sum appearing on the right hand side of Eq.~(\ref{Eq:IHadamard}) is maximized by an expression of the form $M'f(t)$ under constraints $0 < t < 1/e^2$ and $M't \le 1$ to guarantee that the sum of the probabilities does not exceed one.

In the case of orthogonal words based on the Hadamard construction one has $M'\le M$. In this case
a straightforward calculation shows that the mutual information is maximized by one of two strategies.
The first one uses $M'=M$ and $t=1/M$. The resulting bound on the PIE has the simple form $\log M$. This value can be achieved by feeding Hadamard words into a linear optical circuit which maps them onto the PPM format subsequently detected by single photon detection \cite{Guha2011, Banaszek2017}. The second strategy is to use $M'=M-1$ Hadamard sequences with probabilities equal to $t^\ast = 1/e^3$ which maximize individual terms in Eq.~(\ref{Eq:IHadamard}). The residual probability $1-(M-1)t^\ast$ should be divided equally between the remaining Hamadard sequence and its phase flipped version that need to be detected by a type-$\mathscr{Z}$ measurement. The resulting PIE is $2+(M-1)/e^3$. The first strategy is optimal for $M\ge 16$, while the second one offers higher PIE for $M=2,4,8,12$. Its special case for $M=2$ is the measurement described in Sec.~\ref{Subsec:2copy}. These observations are consistent with the results presented in Ref.~\cite{Klimek2016}. It should be kept in mind that the limiting value of PIE is attained when $M\bar{n} \ll 1$ and terms of the order $(M\bar{n})^2$ and higher can be neglected when calculating the photocount probability. When selecting the communication protocol for a given $\bar{n}$ it may be optimal to choose a finite $M$ \cite{Jarzyna2015}.

\section{Conclusions}
\label{sec:conclusions}

We presented a systematic expansion of mutual information for a given ensemble of quantum states and a fixed measurement in the cost parameter characterizing the input ensemble, assuming that with vanishing cost all states converge to one zero-cost state. In the case of optical communication with coherent states, the cost is naturally quantified in terms of the average photon number and the zero-cost state is the vacuum state. Mutual information can be written as a sum of contributions produced by individual measurement outcomes. The expansion depends on whether the measurement operator corresponding to a given outcome detects the zero-cost state or not. In the optical domain, these two classes of operators can be interpreted as photon counting and phase-sensitive detection such as homodyning. The developed formalism substantially helps in deriving bounds on the attainable mutual information in the low-cost limit, reducing the dimensionality of the relevant Hilbert space to zero- and one-photon Fock states and mapping the geometry of the ensemble onto a set of fixed state vectors that are independent of the cost parameter.

In the case of individual measurements on a coherent state ensemble with a vanishing mean photon number, we derived bounds on the attainable photon information efficiency that go beyond those obtained assuming conventional detection techniques. Furthermore, superadditivity of accessible information has been analysed for words composed of the BPSK alphabet of coherent states. The mathematical structure of the derived general bound has been used to identify collective measurements that offer performance beyond individual detection.
In the case when single photon states characterizing the words in the low-cost limit are orthogonal one can calculate an analytical value for the asymptotic maximum PIE and the design of the collective measurement is apparent. The remaining cases in which one cannot assume orthogonality are much more involved and in general one need to resort to numerical optimization. We presented experimental schemes based on linear optics, homodyne detection, and photon counting which in the low-cost limit offer superadditive enhancement over the single-symbol measurements equal to $2.49\%$ and $3.40\%$ respectively for two- and three-symbol measurements.
Their performance for finite mean photon number has been analyzed using as a benchmark the minimum-error measurement on single symbols, which is known to maximize the mutual information for individual detection.
Interestingly, further optimization of collective detection requires a careful adjustment of the measurement to the value of cost parameter.

The developed mathematical formalism can find applications not only in standard classical communication but also in private communication, where the private communication rate can be expressed by the difference of the mutual informations between sender and receiver and the eavesdropper \cite{Wilde2013,Devetak2005}.
An extension of our methods in line with \cite{Czajkowski2017} to tackle von Neumann or Renyi entropies may allow to analyze low-cost limits of various quantum information quantities like Holevo information \cite{Holevo1973}, coherent information \cite{Schumacher1996, Lloyd1997} or other quantities, relevant in quantum key distribution.

\section*{Acknowledgements}

We acknowledge insightful discussions with F. E. Becerra, R. Garc\'{\i}a-Patr\'{o}n, S. Guha and M. G. A. Paris.
This work is part of the project ``Quantum Optical Communication Systems'' carried out within the TEAM programme of the Foundation for Polish Science co-financed by the European Union under the European Regional Development Fund.

\appendix

\section{Photon information efficiency for displaced thermal states}
\label{Sec:thermalNoise}

The treatment of individual measurements performed on coherent states presented in section~\ref{Sec:Individual} can be generalized to the case of displaced thermal states
\begin{align}
		\hat{\varrho}_j = \hat{D}(\gamma_j \zeta) \hat{\varrho}_{\textrm{th}} \hat{D}^{\dagger}(\gamma_j \zeta),
\end{align}
where $\hat{D}(\gamma_j \zeta)$ is the displacement operator and
\begin{align}
	\hat{\varrho}_{\textrm{th}} = \frac{1}{1+n_b} \sum_{n=0}^{\infty} \left( \frac{n_b}{1+n_b} \right )^n \ketbra{n}{n}
	\label{eq:thermal}
\end{align}
is a thermal state with the mean photon number $n_b$, which plays the role of the zero-cost state $\hat{\varrho}^{(0)} = \hat{\varrho}_{\textrm{th}}$. Such states are produced e.g.\ as a result of propagation of coherent states through a phase-invariant Gaussian channel with excess noise.

In the present case the zero-cost state is mixed and has support on the entire bosonic Hilbert space. Thus all POVM operators $\hat Q_r$ need to be type-$\mathscr{Z}$.
For a displaced thermal state, the SLD difference $\hat{D}_j$  at $\zeta = 0$ defined in Eq.~(\ref{Eq:Ldef}) reads
\begin{align}
	\hat{D}_j = \frac{2}{1+2 n_b} \left[ \left(\gamma_j -\gamma_{\textrm{ens}}\right) \hat{a}^{\dagger} + \left(\gamma_j^* -\gamma_{\textrm{ens}}^* \right) \hat{a} \right],
	 \label{eq:SLD_thermal}
\end{align}
where $\hat{a}^{\dagger}$ and $\hat{a}$ denote respectively the creation and the annilation operators of the bosonic mode.

Inserting Eq.~(\ref{eq:SLD_thermal}) into Eq.~(\ref{Eq:Ir<=SLD}) yields an upper bound on mutual information ${\sf I}$ in the low cost limit in the form
\begin{align}
	{\sf I} &= \sum_r {\sf I}_r \le  \frac{\zeta^2}{2} \sum_j p_j \Tr[ \hat{D}_j^2 \hat{\varrho}^{(0)} ] \nonumber \\
	&= \frac{2 \zeta^2}{(1+2 n_b)^2} \sum_j p_j \abs{\gamma_j -\gamma_{ens}}^2 \Tr [\hat{a}^\dagger
\hat{a} \varrho^{(0)}+ \hat{a} \hat{a}^\dagger \varrho^{(0)} ] \nonumber\\
	&= \frac{2 \bar n}{1+2 n_b} \label{eq:mi_thermal}
\end{align}
where we have used $\Tr[(\hat{a}^{\dagger})^2 \rho^{(0)}] = 0$ and $\Tr[\hat{a}^2 \rho^{(0)}] = 0$ as well as the definition of the mean photon number from Eq.~(\ref{Eq:nbar=}). Note that Eq.~(\ref{eq:mi_thermal}) is identical with the first order expansion of the Shannon-Hartley formula \cite{Shannon1949} for the scenario with excess noise
\begin{align}
{\sf I}_{\textrm{SH}} = \frac{1}{2} \log\left(1+\frac{4 \bar n}{1+ 2 n_b}\right) \; .	
\end{align}

\section{Homodyne measurement schemes}
\label{Sec:HomodyneScheme}

The statistics of outcomes for shot-noise limited homodyne detection of the real quadrature $x$ on a coherent state with a complex amplitude $\zeta$ is given by a Gaussian probability distribution
\begin{equation}\label{eq:prob_homo}
p_{\text{hom}}(x|\zeta) = \frac{1}{\sqrt{\pi}} \exp[-(x-\sqrt{2} \text{Re}\, \zeta)^2].
\end{equation}
Mutual information for homodyne detection of the BPSK ensemble of coherent states with real amplitudes $\ket{\zeta}$ and $\ket{-\zeta}$ can be obtained by plugging Eq.~(\ref{eq:prob_homo}) into Eq.~(\ref{eq:I}) and (\ref{Eq:Ir})
\begin{align}
{\sf I}_{\text{BPSK/hom}} = 4\zeta^2  - \frac{e^{-2\zeta^2}}{\sqrt{\pi}} \int_{-\infty}^{\infty} dx \, e^{-x^2} \cosh(2\sqrt{2}x\zeta) \log [ \cosh(2\sqrt{2}x\zeta)],
\end{align}
with the summation over results exchanged to integration. Expanding all the functions in the second term up to the fourth order in $\zeta$ yields
\begin{align}
	{\sf I}_{\text{BPSK/hom}} \approx
4\zeta^2  - \frac{1-2\zeta^2+2\zeta^4}{\sqrt{\pi}} \int_{-\infty}^{\infty}  dx \, e^{-x^2} \left( 4 x^2 \zeta^2 + \frac{32}{3} x^4 \zeta^4 \right)
	\approx 2 \zeta^2 - 4 \zeta^4.
\end{align}
The rightmost expression coincides with the power series expansion of the Shannon-Hartley limit given in Eq.~(\ref{Eq:IindExpansion}).

The statistics outcomes on the single photon detector (SPD) deployed in the measurement scheme discussed in Sec.~\ref{Subsec:2copy} and depicted in Fig.~\ref{Fig:2copy_detect}
with an optional displacement $\beta$ is given by
\begin{equation}
p_{\text{SPD}}(k|\zeta') = \begin{cases}
	e^{-|\zeta'+\beta|^2} , & k = 0 \\
	1 - e^{-|\zeta'+\beta|^2} , & k =1
\end{cases}
\end{equation}
where $k=0,1$ indicates respectively if any photons were registered or not and $\zeta'$ is the field amplitude before the displacement operation.
The probabilities of possible outcomes in the two-symbol measurement design of Fig.~\ref{Fig:2copy_detect} are then given by
\begin{align}
p_{x,k|{\textbf j}} = p_{\text{hom}}\left(x\Big|[(-1)^{j_1}+(-1)^{j_2}] \frac{\zeta}{\sqrt{2}}\right) p_{\text{SPD}}\left(k\Big |[(-1)^{j_1}-(-1)^{j_2}] \frac{\zeta}{\sqrt{2}}\right)
\end{align}
The mutual information can be written as
\begin{align}
\label{Eq:I2App}
	{\sf I}_2 &=  H(X,K)-\sum_{\textbf j} p_{\textbf j} \left\{ {\sf H}\left[p_{\text{SPD}}(0|{\textbf j})\right]+h\left[p_{\text{hom}}(x|{\textbf j})\right] \right \}
\end{align}
where $H(X,K)$ denotes the joint entropy of the outcomes at the homodyne detector and the SPD, ${\sf H}$ is the entropy of a binary random variable defined in Eq.~(\ref{Eq:BinEntinNats}) and $h[p(x)] = - \int_{-\infty}^{\infty}  dx \, p(x) \log p(x) $.

In the case of a pure SPD measurement without displacement ($\beta = 0$) one can calculate mutual information by expanding $\log p_{x,0}$ appearing in the expression for $H(X,K)$ up the second order in $\zeta$.
Taking the limit of $\zeta^2 = \bar n \rightarrow 0$ in the resulting expression yields
\begin{align}
	{\sf I}_2 \approx 4 \bar n (1-u) -2 \bar n u \log u = 2\bar n [2+f(u)] .
\end{align}
The derived expression is equal to the mutual information of the two-symbol scenario of Eq.~(\ref{Eq:Itwosymbolfinal}) and is maximized by $u = 1/e^3$. The advantage is 2.49\% compared to the individual measurement.

\bibliographystyle{iopart-num}
\bibliography{superaddLib}

\end{document}